\documentclass[12pt]{iopart}
\usepackage{graphicx,iopams}

\usepackage{cite}
\usepackage[none]{hyphenat}
\usepackage[normalem]{ulem}
\expandafter\let\csname equation*\endcsname\relax

\expandafter\let\csname endequation*\endcsname\relax
\usepackage{amsmath}
\usepackage{epstopdf}
\usepackage{soul}
\usepackage[textsize=tiny]{todonotes}

\def\pmb#1{\setbox0=\hbox{#1}
	\kern-.025em\copy0\kern-\wd0 \kern.05em\copy0\kern-\wd0
	\kern-.025em\raise.0433em\box0}

\begin{document}
	
	\title[Phase transitions in one-dimension]{Classical phase transitions in a one-dimensional short-range spin model induced by entropy depletion or complex fields}
	
	\author{P. Sarkanych$^{1,2,3}$, Yu. Holovatch$^{1,3}$, R. Kenna$^{2,3}$}
	
	\address{$^{1}$
			Institute for Condensed Matter Physics, National Acad.
			Sci. of Ukraine, UA--79011 Lviv, Ukraine}
	
	\address{$^{2}$
			Applied Mathematics Research Centre, Coventry University, Coventry
			CV1 5FB, United Kingdom}
	
	\address{$^{3}$
	${\mathbb L}^4$ Collaboration \& Doctoral College for the Statistical Physics of Complex Systems, Leipzig-Lorraine-Lviv-Coventry, Europe}
		
\begin{abstract}
Ising's solution of a classical spin model famously demonstrated the absence of a positive-temperature phase transition in one-dimensional equilibrium systems with short-range interactions. 
No-go arguments established that the energy cost to insert domain walls in such systems is outweighed by entropy excess so that symmetry cannot be spontaneously broken. 
An archetypal way around the no-go theorems is to augment interaction energy by increasing the range of interaction. 
Here we introduce new ways around the no-go theorems by investigating entropy depletion instead. We implement this for the Potts model with invisible states.
Because spins in such a state do not interact with their surroundings, they contribute to the entropy but not the interaction energy of the system.
Reducing the number of invisible states to a negative value decreases the entropy by an amount sufficient to induce a positive-temperature classical phase transition.
This approach is complementary to the long-range interaction mechanism.
Alternatively, subjecting positive numbers of invisible states to imaginary or complex fields can trigger such a phase transition.
We also discuss potential physical realisability  of such systems.
\end{abstract}

 \submitto{\JPA}
	
\eads{\mailto{sarkanyp@coventry.ac.uk}, \mailto{hol@icmp.lviv.ua}, \mailto{r.kenna@coventry.ac.uk}}

\section{Introduction}
\label{I} 

As is widely known, in his famous 1925 paper \cite{Ising}, and following a suggestion by Wilhelm Lenz, Ernst Ising sought a positive-temperature  phase transition in a one-dimensional (1D) classical equilibrium system with short-range interactions \cite{Ising17}. 
To some disappointment \cite{Rowlinson}, there was none. 
This was the start of a vast amount of literature on the statistical mechanics of critical phenomena, including a number of studies on why it is impossible to have a phase transition in such systems \cite{Landau,Hove,Ruelle1969,CuSa}. 
The lower critical dimension is now defined as that below which a phase transition cannot occur at positive temperature
and at least two physical dimensions are required for many short-range classical equilibrium models. 

Landau and Lifschitz gave heuristic arguments suggesting that entropic excesses prevent phase transitions below the upper critical dimension\cite{Landau}; 
van Hove's approach was based on proofs of analyticity of the transfer matrix eigenvalues and free energy \cite{Hove}; 
Ruelle extended this giving rigorous theorems~\cite{Ruelle1969} and, more recently, 
Cuesta and S{\'{a}}nchez \cite{CuSa} presented more general results about the non-existence of phase transitions in 1D short-ranged systems. 
For such classical, equilibrium models with short range interactions in 1D second-order phase-transition type phenomena can only occur at zero temperature. 
The essence of early no-go arguments is that there is an entropy excess in 1D systems relative to interaction energy so that the delicate balance that gives a phase transition is not achieved.
The role played by domain walls was further investigated in Ref.~\cite{Theo08}.

To escape the limitations of no-go theorems, interactions with sufficiently long range can be introduced  \cite{Anderson,Dyson71,FrSp82}. 
Another way out is provided by non-equilibrium systems~\cite{Evans} and
further  exceptions are discussed in Ref.~\cite{CuSa}.
In Ref.~\cite{Sarkanych2017} we introduced a new way around the no-go theorems involving models with a negative number of invisible states or complex fields acting on them.
Here we develop this circumvention through a Lee-Yang-zero analysis \cite{LY}, an approach described as fundamental to the theory of phase transitions \cite{fund}.
We also discuss potential physical realisations in real-world systems.

The Potts model with invisible states was introduced nearly a decade ago in Refs.~\cite{Tamura2010,Tanaka2011}. 
It differs from the ordinary Potts model \cite{Potts1952} in that spins in an invisible state do not interact with their neighbours but they do contribute to entropy. 
The corresponding Hamiltonian takes the form
\begin{equation}
H=-\sum_{<i,j>}\delta _{s_i,s_j}\sum_{\alpha=1}^q \delta
_{s_i,\alpha} {\delta _{s_j,\alpha}},
\label{one}
\end{equation}
where $q$ and $r$ are the number of visible and invisible states respectively, $s_i=1,\ldots,q,q+1,\ldots,q+r$ is the Potts variable and $\delta$ is Kronecker delta symbol. 
The first sum in Eq.~(\ref{one}) is taken over all distinct pairs of interacting particles, and the second sum requires both of the interacting spins to be in the same visible state. 
We henceforth use the term ``$(q,r)$-state Potts model'' for systems with $q$ visible and $r$ invisible states.

The usual concept of universality means that critical behaviour is determined by dimensionality, the range of the interaction   and the symmetries of the system. 
Although the number of invisible states  $r$ does not change any of these properties, it was shown to control the order of the phase transition in Refs.~\cite{Tamura2010,Tanaka2011,Krasnytska2016}. 
For example although the two-dimensional $(2,0)-$state Potts model (which is the ordinary Ising model) is the archetypal example of a continuous phase transition, the model with $(2,30)$ states undergoes a first-order transition.

We confirm that the Potts model with a positive number of invisible states adheres to the no-go theorems in one dimension in that the only possibility for a phase transition is at zero temperature.
As a positive number of invisible states delivers additional entropy to the system, further preventing a positive-temperature phase transition, one may consider that a negative number corresponds to a depletion of entropy.
We show that if the number of invisible states is negative and large enough by  absolute value the phase transition can be shifted to a positive value of temperature. 
The same phenomenon can be achieved through introducing a complex external field which acts on the invisible states.
Although some of these concepts are unphysical in and of themselves, they can be linked with physicality in a number of interesting ways.

Long thought to be entirely mathematical constructs whose roles, while important, are restricted only to fundamental theories that underlie phase transitions, the notion of complex magnetic fields has recently gained physical traction too.
Following on from an earlier theoretical proposal \cite{Wei2012}, Peng et. al. have shown that complex fields associated with a spin bath are related to the quantum coherence of a probe spin coupled to the bath~\cite{Peng2015}. 
The results demonstrate that the times at which  quantum coherence reaches zero are equivalent to the complex values of magnetic fields at which the partition function vanishes: i.e., Lee-Yang zeros \cite{LY}.
The zeros method is considered of fundamental importance in understanding phase transitions 
\cite{fund} and a  powerful tool to analyse  critical behaviour \cite{Bena2005}.
The Lee-Yang zeros give direct access to the partition function itself and as such give important information on the nature of phase transitions.
Very recently exact results on the classical antiferromagnetic Ising chain in a magnetic field
showed an infinite cascade of thermal phase transitions, the origins of which were traced to the lines of the Lee-Yang zeros, opening a way to relate  to observable and potentially measurable quantities \cite{TiCh17}.

Low-dimensional models are of continued theoretical and physical interest \cite{CuSa}.
A new combinatorial approach was used to solve Ising's model in Ref.~\cite{Seth2017} and it was suggested that the method could be applied to the 2D problem.
The first  experimental  verification of Onsager's 1943 solution of the 2D Ising model \cite{Onsager} is also a very recent development, offering a ``promising candidate for numerous applications'' \cite{LeLe16}.
This recent experimental advance inspired the question ``are we going to see a transition \dots in a chain of \dots molecules (1D)'' \cite{Press}.
This is the question addressed in this paper.

For these reasons, we analyse 1D models with invisible states using Lee-Yang zeros.
There have been other approaches to access phase transitions in 1D models.
Following on from suggestions by Anderson \cite{Anderson}, Dyson 
\cite{Dyson71} proved that systems with long range order can have  positive-temperature phase transitions and
Fr{\"{o}}hlich and Spencer proved the existence of a spontaneous magnetization at positive low
temperature for the one-dimensional Ising Model with long-range interactions.
Cuesta and S{\'{a}}nchez  gave three further examples of phase transitions, both of purely academic interest
and with importance for phenomena such as surface growth and DNA denaturation~\cite{CuSa}. 
For quantum phase transitions the critical dimensionality is also reduced relative to the corresponding
classical transition \cite{Suzuki1976}.
Some chemical compounds are well described by quasi one dimensional models 
 \cite{Rice1993,Rice1997,Azuma1994,Roger1983,Hovhannisyan2009,Ananikian}.
Here we are interested in pure 1D equilibrium models which are both classical and short-range.

The Potts model with invisible states describes a number of models of physical interest. 
Notably, the $(1,r)$-state case is equivalent to the Ising model in a temperature dependent field and can be mapped to the Zimm-Bregg model for the helix-coil transition \cite{Badasyan2010}. 
The long-range extension of the $(1,r)-$state model possesses a reentrant phase transition and is in a good agreement with experimental observations for polymer transitions\cite{Badasyan2011}. 
The $(2,r)-$state Potts model without external fields is equivalent to the Blume-Emery-Grifiths  model \cite{Tamura2010,Johnston2013,Ananikian2013}.
The general $q$ and $r$ case can be interpreted as a diluted Potts model\cite{Tamura2010,Krasnytska2016}.

In Section \ref{II} we present the exact solution for the 1D Potts model with invisible states using the {well-known transfer matrix approach \cite{Baxter1982,Katsura1972,Kim2000,Shrock1997}. 
Some results from such well-established material were  presented in the letter \cite{Sarkanych2017} where the focus was on  Fisher-zeros \cite{Fisher1965} (zeros in the complex temperature plane) and we elaborate on these in \ref{Fisher}.
Here our focus is on Lee-Yang zeros (zeros in the complex magnetic-field plane).
We present a duality relation between field and temperature in \ref{Duality}.
In Section~\ref{III} we investigate the Lee-Yang zeros for the ordinary Potts model and for its counterpart with a positive number of invisible states.  Singular behaviour of the Lee-Yang zeros is discussed in \ref{Singular}.
In Section~\ref{IV} we relax the constraints involved in conventional studies by allowing the number of invisible states to be negative and/or the magnetic fields to be complex. 
These enable positive-temperature phase transitions to be achieved --- of the type sought by Ising nearly 100 years ago.
We draw our conclusions in Section~\ref{V}.

\section{Potts model with invisible states}
\label{II} 

We consider the Potts model with invisible states (the $(q,r)-$Potts model) with nearest-neighbour interactions 
on a 1D chain of $N$ spins with periodic boundary conditions. 
The partition function is
\begin{equation}
Z=\sum_{s}\exp\left(-\beta H_{(q,r)}\right),
\end{equation}
where $\sum\limits_s$ denotes the sum over all possible spin configurations.
With periodic boundary conditions the Hamiltonian can be rewritten 
as a sum of terms representing one bond each, namely
\begin{equation}
H_{(q,r)}=\sum_{i} H_{i},\quad \text{where}\quad H_{i}=-\delta_{s_i,s_{i+1}}\sum_{\alpha=1}^q\delta_{s_i,\alpha}-h_1\delta_{s_i,1}-h_2\delta_{s_i,q+1},
	\label{Jub}
\end{equation}
where  the variable $i$ spans the $N$ sites of the chain, $s_i=1, \ldots ,q,q+1, \ldots ,q+r$ is a Potts variable and $h_1$ and $h_2$ are two ordering fields acting on the first visible and first invisible states respectively,
so that
\begin{equation}
Z=\sum_{s}\prod_{i}\exp\left( -\beta H_i\right).
\end{equation}

The final term in Eq.~(\ref{Jub}) selects only one of the $r$ invisible states as interacting with the external field $h_2$.
As such, it contributes to the energy if $h_2 \ne 0$.
The $r-1$ remaining identical invisible states contribute only to the entropy, as do all invisible states if $h_2$ vanishes. 
This means that different microscopic configurations could be understood as the same macroscopic configuration. 
In terms of the partition function the effect  is to multiply some of the terms by $(r-1)$. 
Similarly, as was done in Ref.~\cite{Tamura2010}, we can collect all invisible states into a single one with appropriate weight and consider the equivalent Hamiltonian of a diluted Potts model:
\begin{equation}
H_{(q,r)}^{eq}=-\sum_{i} \delta_{\sigma_i,\sigma_{i+1}}\sum_{\alpha=1}^q\delta_{\sigma_i,\alpha}-h_1\sum_i \delta_{\sigma_i,1}-h_2\sum_i\delta_{\sigma_i,q+1}-T\ln(r-1)\sum_i\delta_{\sigma_i,q+2}, 
\label{Href}
\end{equation}  
where $\sigma_i=1, \ldots ,q,q+1,q+2$ is a new Potts variable and all (except the one along the field $h_2$) the invisible states are gathered into one with the appropriate weight. The Hamiltonian (\ref{Href}) is, of course, different to that in Eq. (\ref{Jub}). But the corresponding partition functions are the same.

\subsection{Transfer matrix}
\label{TM}

To develop the formalism to solve the model (\ref{Jub}) exactly, we define the transfer matrix 
${{\mathbf{T}=\mathbf{T}}(s_i,s_j)}$ as \cite{Baxter1982,Katsura1972,Kim2000,Shrock1997}
\begin{equation}
{{\mathbf{T}}(s_i,s_j)=\exp\left[\beta(\delta_{s_i,s_j}\sum_{\alpha=1}^q\delta_{s_i,\alpha} + h_1\delta_{s_i,1} + h_2\delta_{s_i,q+1})\right]},
\end{equation}
so that, in the explicit form of a $(q+r) \times (q+r)$ matrix, 
\begin{equation}
\label{transfermatrix}
\mathbf{T}= 
 \begin{pmatrix}
 yz_1 & 1 & 1 & \cdots & 1 & z_2 & 1 & \cdots & 1 \\
 z_1  & y & 1 & \cdots & 1 & z_2 & 1 & \cdots & 1 \\
 z_1  & 1 & y & \cdots & 1 & z_2 & 1 & \cdots & 1 \\
 \vdots  & \vdots  & \vdots &\ddots & \vdots & \vdots& \vdots& \ddots & \vdots \\
 z_1 & 1 & 1 & \cdots & y & z_2 & 1 & \cdots & 1 \\
 z_1 & 1 & 1 & \cdots & 1 & z_2 & 1 & \cdots & 1\\
 \vdots &\vdots &\vdots &\ddots &\vdots &\vdots &\vdots &\ddots &\vdots \\
 z_1 & 1 & 1 & \cdots & 1 & z_2 & 1 & \cdots & 1
 \end{pmatrix}
\end{equation}
where the columns are given by the values of $s_i$ and the rows are given by the values of $s_{i+1}$
and where temperature and field dependencies have been absorbed into the variables 
\begin{equation}
y=e^\beta, \quad
z_1=e^{\beta h_1}, \quad 
z_2=e^{\beta h_2}.
\end{equation}
On par with the temperature variable $y$ in the text we will be using another variable $t=y^{-1}=e^{-\beta}$. It is useful since the infinite range $0\leq T< \infty$ converges to the finite region $0\leq t\leq 1$.
 
The partition function can then be recast as
\begin{equation}
\label{partfunc}
Z=\prod_{i=1}^{i=N}\sum_{\{s_i\}}\mathbf{T}(s_i,s_{i+1})=\Tr \mathbf{T}^N=\sum_{i}\lambda_i^N,
\end{equation}
where $\lambda_i$ are the eigenvalues of $\mathbf{T}$.

Some of the eigenvalues can be found using the symmetry of the transfer matrix. It is easy to show that matrix (\ref{transfermatrix}) has five different eigenvalues. 
On the one hand, because the final $r$ columns of the matrix are proportional, one eigenvalue is zero and is $r-1$ times degenerate. 
On the other hand, because $(q-1)$ elements of the main diagonal are equal to $y$,  choosing $\lambda=y-1$ leads to $q-2$ linearly independent eigenvectors.  This leaves only three unknown  eigenvalues. They can be found using invariant permutations. This approach leads to the equation for the three remaining eigenvalues:
\begin{equation}
(r-1-\lambda +z_2)(yz_1-\lambda-z_1)(y-\lambda-1)-\lambda z_1(y-\lambda-1)-(q-1)(yz_1-\lambda-z_1)\lambda=0.
\label{eigmain}
\end{equation}
This is an equation of third power and  can, therefore, be solved exactly. 
Since the partition function (\ref{partfunc}) is defined by eigenvalues and all the $\lambda$'s have been found, the problem is solved exactly \cite{Sarkanych2017}.

\subsection{Partition function zeros}
\label{2.2}

Critical behaviour of equilibrium systems can be extracted from the partition function. 
In our case, the latter  is described by the eigenvalues of the transfer matrix (\ref{partfunc})
and, since we have shown that they can all  be found explicitly, 
the critical properties of the Potts model with invisible states can, in principle, also be found explicitly. 
This allows us to access the Lee-Yang zeros in complex magnetic field \cite{LY}.
For completeness, we also discuss the Fisher zeros \cite{Fisher1965} in the complex temperature planes in \ref{Fisher}.

The standard approach is to label the eigenvalues of the transfer matrix in such a way that they are ordered in magnitudes;
$ |\lambda_1|\leq|\lambda_2|\leq|\lambda_3|\leq \ldots$ .
The partition function zeros are then found using the condition that (at least) two eigenvalues are largest by modulus \cite{Fisher1980} 
\begin{equation}
\label{condmodulus}
|\lambda_1|=|\lambda_2| \, .
\end{equation}
Since the partition function is analysed in the complex ($T$ or $h$) plane, the eigenvalues are complex as well. Therefore condition (\ref{condmodulus}) can be written as
\begin{equation}
\label{eqlambda}
\lambda_2=\lambda_1 e^{i\phi}\, .
\end{equation}

From Eq.~(\ref{partfunc}), the partition function is a sum of eigenvalues to the power $N$. 
In our case the eigenvalue $\lambda=0$ makes no contribution so that the partition function takes the form $Z=\lambda_1^N+\lambda_2^N+\lambda_3^N+\lambda_4^N$. 
Of these four eigenvalues,  three are roots of the polynomial (\ref{eigmain}) and the fourth  equals $y-1$. 
Taking into account the orders of magnitude of the eigenvalues. 
the partition function may be rewritten in such a way as to single out the main terms;
\begin{equation}
\label{partfunclambda}
Z=\lambda_1^N
\left[{
1+e^{iN\phi}+\left(\frac{\lambda_3}{\lambda_1}\right)^N+\left(\frac{\lambda_4}{\lambda_1}\right)^N
}\right]\,.
\end{equation}
In the thermodynamic limit $N \rightarrow \infty$ only the leading two terms in the expression in parentheses on the 
right-hand side of Eq.~(\ref{partfunclambda}) contribute so that we obtain the phase $\phi$ given by
\begin{equation}
\label{phicond}
1+e^{iN\phi}=0 \quad {\mbox{or}} \quad \phi=\frac{2k-1}{N}\pi, \quad k=1\ldots N \, .
\end{equation}

Therefore the coordinates of the partition function zeros are found solving Eq.~(\ref{eqlambda}) with the phase given by Eq.~(\ref{phicond}). 
This method is appropriate when all the eigenvalues are given explicitly. 
However, when they are given as the roots of the polynomial, 
 one can use the method suggested in Ref.~\cite{Ghulghazaryan2007} for  models with three non-zero eigenvalues or that put forward in Ref.~\cite{Hovhannisyan2009} adapted for  models with four non-zero eigenvalues.

Following this method, the four eigenvalues of the transfer matrix are presented as the roots of the polynomial of  fourth order \footnote{ Since one does not know in advance which eigenvalue is the maximum one, all four eigenvalues have to be considered.}. In the most general case it has the form
\begin{equation}
\label{fourthorderpol}
\lambda^4+a_3\lambda^3+a_2\lambda^2+a_1\lambda+a_0=0 \,.
\end{equation}
For the Potts model with invisible states Eq.~(\ref{fourthorderpol}) is obtained by multiplying Eq.~(\ref{eigmain}) by $[\lambda-(y-1)]$. The corresponding coefficients have the form
\begin{eqnarray}
&&\nonumber a_0=(y-1)^3 z_1 (r+z_2-1)\,; \\
&&\label{coefficients} a_1=-(y-1)^2 (z_1 (q+y-1)+(2 z_1+1) (r+z_2-1))\,; \\ 
&&\nonumber a_2=(y-1) ((z_1+1) (q+y-1)+(z_1+2)(r+z_2-1)+y z-1)\,; \\ 
&&\nonumber a_3=-q-r-y z_1-2 y-z_2+4\,.
\end{eqnarray} 
The goal of the method is to obtain a $\lambda-$independent equation linking together temperature, fields and other model parameters. 
To derive this we use four equations from Vieta's theorem together with the condition (\ref{eqlambda}). Excluding all eigenvalues from these five equations we obtain
\begin{equation}
 F(q, r, z_1, z_2, y)=F_1 F_2(f_1+f_2+f_3+f_4) ,
\label{largeeq}
\end{equation}
where
\[
 {F_1}=8 a_2 \cos ^2\left(\frac{\phi }{2}\right) \cos (\phi )-a_3^2 [2 \cos (\phi )+1];\\
\]
\[
F_2=4 a_1 [2 \cos (\phi )+1] \left[\cos \left(\frac{\phi }{2}\right)+\cos \left(\frac{3 \phi }{2}\right)\right]^2-32 a_2 a_3 \cos ^4\left(\frac{\phi }{2}\right) \cos{\phi} 
 a_3^3 [2 \cos (\phi )+1]^2;
\]
and
\begin{eqnarray*}
\hspace{-2.5cm}
f_1  & = & 16 a_0^3 \left[\cos \left(\frac{\phi }{2}\right)+\cos \left(\frac{3 \phi }{2}\right)\right]^4;
\\
\hspace{-2.5cm}
 f_2 & = & a_1^2\bigg[a_2^2 \left(a_3^2-2 a_2 (\cos (\phi )+1)\right)-a_1^2(1+2 \cos (\phi ))^3+
\\
\hspace{-2.5cm} & &  \bigg. 2 a_1 a_3 \bigg(a_2 (5 \cos (\phi )+\cos (2 \phi )+3)-a_3^2 (1+\cos (\phi ))\bigg)\bigg];\\
\hspace{-2.5cm} f_3 & = & -2a_0\left[a_2^3 \bigg(8 a_2 \cos ^4\left(\frac{\phi }{2}\right)-a_3^2 (\cos (\phi )+1)\bigg)+\right.
\\
\hspace{-2.5cm} && a_1^2 \bigg(4 a_2 \cos ^2\left(\frac{\phi }{2}\right) (7 \cos (\phi )+5 \cos (2 \phi )+\cos (3 \phi )+5)- a_3^2 (2 \cos (\phi )+\cos (2 \phi ))\bigg)+ 
\\
\hspace{-2.5cm} && \left.a_1 a_2 a_3 \bigg(-4 a_2 \cos ^2\left(\frac{\phi }{2}\right) (6 \cos (\phi )+\cos (2 \phi )+3)+a_3^2 (5 \cos (\phi )+\cos (2 \phi )+3)\bigg)\right]; 
\\
\hspace{-2.5cm} f_4 & = & a_0^2 \bigg[128 a_2^2 \cos ^4\left(\frac{\phi }{2}\right) \cos ^2(\phi )+a_3^4(2 \cos (\phi )+1)^3-\bigg.
\\
\hspace{-2.5cm} && 8 a_2 a_3^2 \cos ^2\left(\frac{\phi }{2}\right) \bigg(7 \cos (\phi )+5 \cos (2 \phi )+\cos (3 \phi )+5\bigg)+
\\
\hspace{-2.5cm} && 
\bigg.
			       8 a_1 a_3 
						 \left(
						       \cos 
									     \left(
											       \frac{\phi }{2}
											 \right)
											 +
									 \cos 
									     \left(
											        \frac{3 \phi }{2}
											 \right)
						\right)^2 
						(2 \cos (\phi )+\cos (2 \phi )+3)
\bigg].
\end{eqnarray*}

In  Eq.~(\ref{largeeq}) $\phi$ takes discrete values according to Eq.~(\ref{phicond}). 
All the roots of Eq.~(\ref{largeeq}) provide the values of parameters when two eigenvalues are equal by modulus, but not all of them are actual partition function zeros. 
Actual zeros are characterised by the condition that two largest eigenvalues are equal by modulus.

We analyse the zeros of the partition function in the plane of complex magnetic field (Lee-Yang zeros). 
According to the Lee-Yang theorem, for the ferromagnetic Ising model on a $d-$dimensional regular lattice, these zeros
are purely imaginary  \cite{LY}. 
This statement can be  generalised to many other models  \cite{Bena2005}. 
Transforming to the complex $z=e^{-\beta h}$ plane, the counterpart zeros lie on an arc of the unit circle. 
Lee-Yang zeros have been called ``protocritical points'' \cite{Fisher1980} because they have the potential to become actual critical points.
The protocritical point at an end of the arc which lies closest to the positive real axis is referred to as the ``Yang-Lee edge'' (henceforth also referred to as the ``edge'')~\cite{Fi78}.
If the temperature is higher than the critical one, the circular arc is open, i.e., it does not cross the positive real axis.
As the temperature is lowered, the arc becomes a circle and the edge pinches the real axis when the critical temperature is reached, precipitating the phase transition.
For temperatures lower than the critical value, the arc is closed into a full circle so that zeros cross the real $z$ axis. 

\section{Lee-Yang zeros for  models with direct physical realisability}
\label{III}

\subsection{Lee-Yang zeros for the ordinary Potts model}
\label{III11}

We first consider the ordinary $q-$state Potts model when the number of invisible states and, correspondingly, the second magnetic field are set to zero ($r=0$ and $h_2=0$).  
This ordinary Potts model is thoroughly investigated so our results can be compared to those previously obtained \cite{Glumac1994}. 
In this case one of the roots of  Eq.~(\ref{eigmain}) becomes $\lambda=y-1$, reducing the number of different eigenvalues to three. 
The remaining two eigenvalues are found as the roots of Eq.~(\ref{eigmain}) and fully recover results obtained in Ref.~\cite{Glumac1994}.
The three eigenvalues are
\begin{equation}
\lambda_{1,2}=\frac12 \left[(y(z_1+1)+q-2)\pm \sqrt{(y(1-z_1)+q-2)^2+(q-1)4z_1}\right], \,
\lambda_3= y-1.
\label{orpr}
\end{equation}
In Ref.~\cite{Fisher1980}  it was shown that the edge can be recovered from the condition that the largest eigenvalues of the transfer matrix are degenerate.  
Two of these eigenvalues in Eq.~(\ref{orpr}) are degenerate when the expression under the square root sign vanishes.

\begin{figure}[t]
	\begin{center}
	\includegraphics[width=0.3\paperwidth]{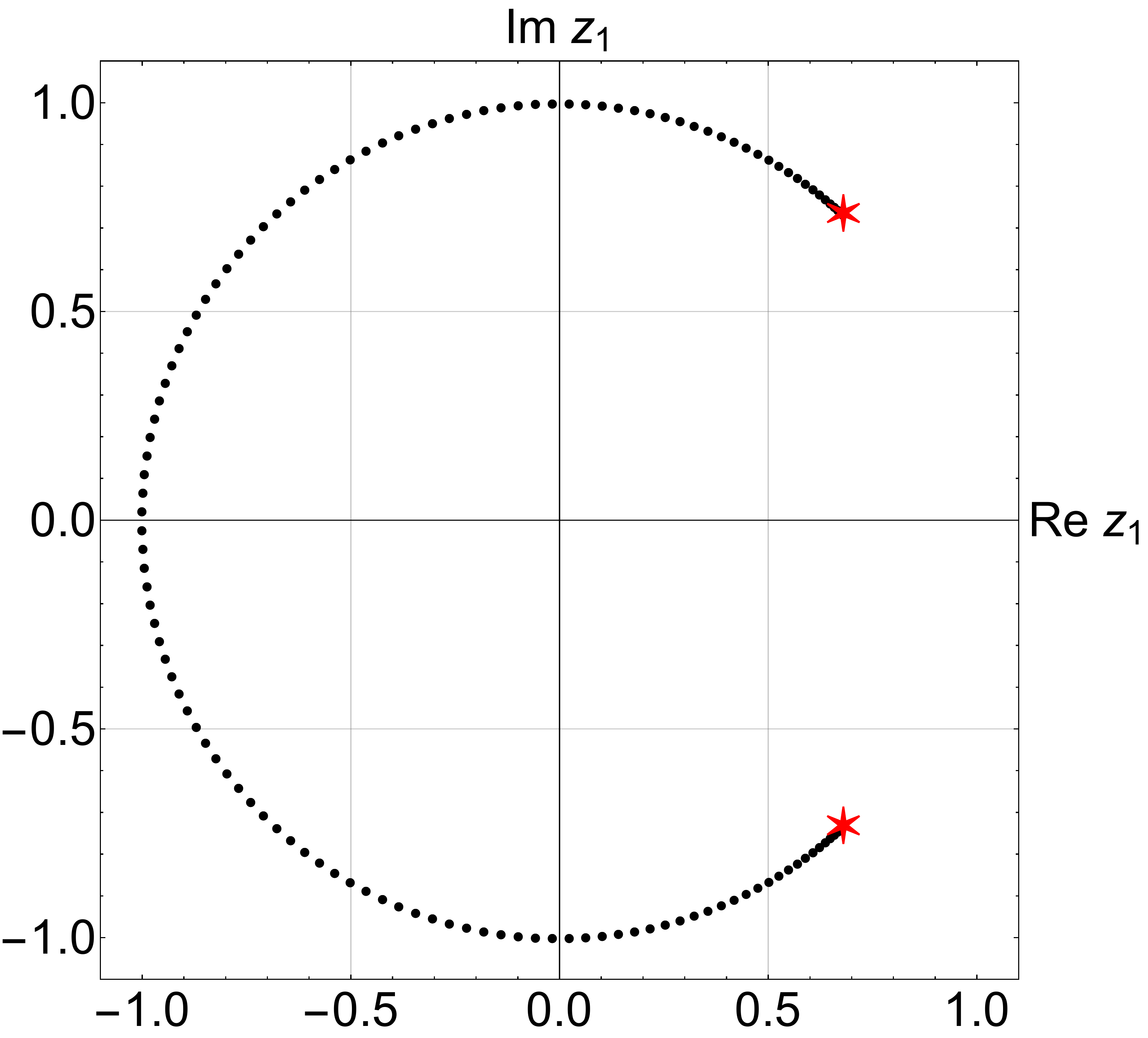}
	\includegraphics[width=0.3\paperwidth]{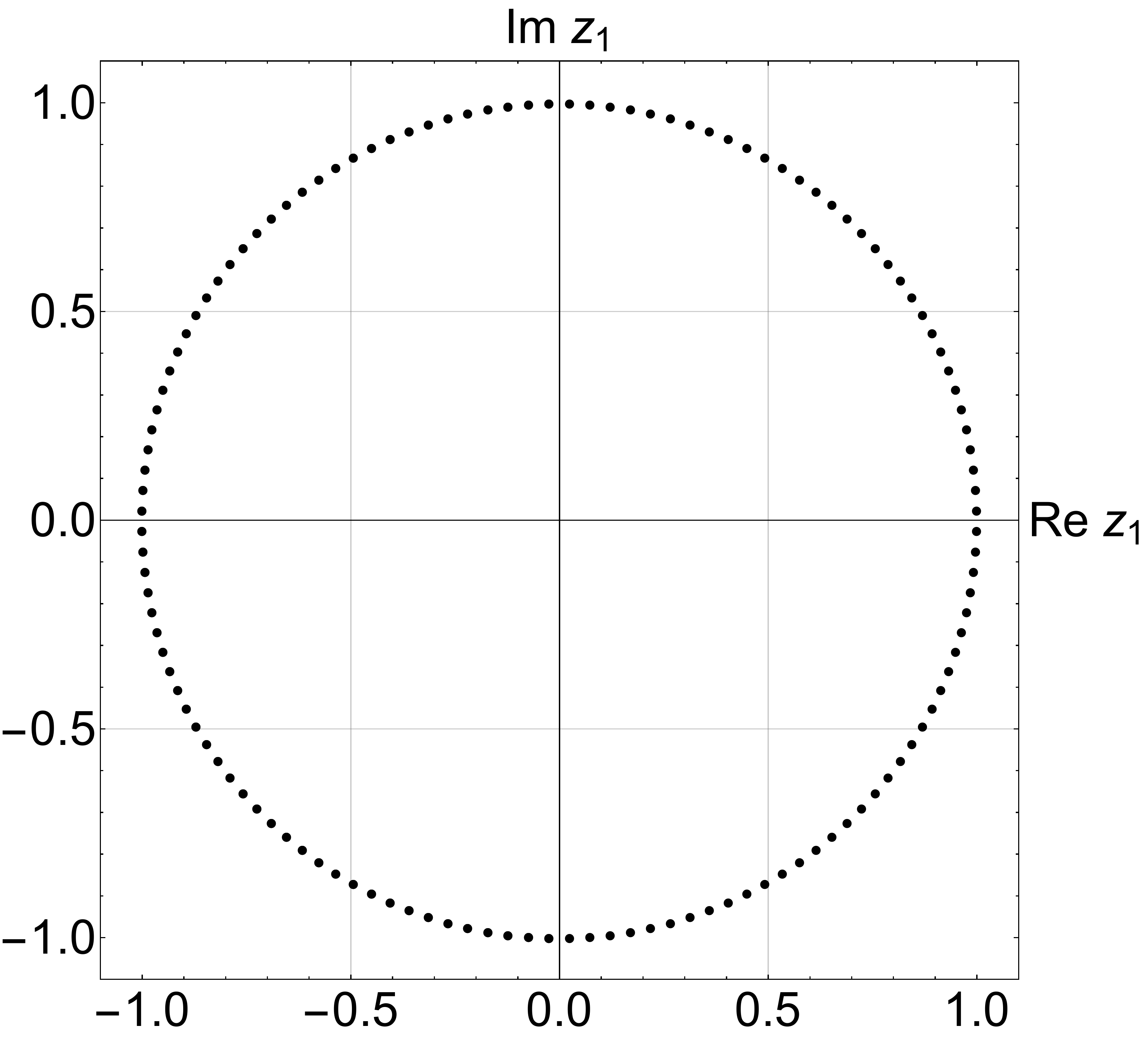}
	\caption{Lee-Yang zeros in the complex ${z_1}=e^{-\beta h_1}$ plane of the 1D Ising model ($q=2, r=0$) for fixed values of $T$. 
  The plots are for finite $N$ but the curves become continuous in the thermodynamic limit.
	The left panel  displays Lee-Yang zeros  with a positive value of the temperature ($T=1.091$).	
  The Yang-Lee edges, represented as large red stars, do not reach the positive real axis indicating no phase transition.
	The right panel  displays zeros  with with vanishing temperature ($T=0$).
	The approach of the zeros to the real axis at ${{z_1}}=1$ at infinite $N$ indicates a zero-field (${{h_1}}=0$ ) phase transition.
 	\label{fig1}}
	\end{center}
\end{figure}

We plot the resulting loci of Lee-Yang zeros of the 1D Ising model ($q=2$) in  Fig.~\ref{fig1}.
The $T>0$ case is illustrated in the left panel. 
There the edge is strictly complex meaning there is no phase transition in the symmetric phase.
As the temperature decreases, the edge approaches the real axis. 
The limiting case of $T=0$ is represented in the right panel, albeit for a finite-size system
(the circle is complete for an infinite chain of sites). 
In the thermodynamic limit the approach of the edge to the critical point ($h=0$ or $z=1$) 
triggers the zero-temperature spontaneous (zero-field) phase transition.
Fig.~\ref{fig1} illustrates the $q=2$ case only, for which the Lee-Yang unit-circle theorem is obeyed.
Altering the number of Potts states alters the loci of zeros (not shown in the plot); 
while they remain circular, their radii are $q$-dependent for positive temperature. 
If $q<2$ the radii of these circles are less than 1 and if $q>2$ the radii exceed 1. 
However, at $T=0$ all Lee-Yang arcs close into circles and cross the real axis at ${\rm Re}\,z_1=1$.

\begin{figure}[t]
	\begin{center}
    \includegraphics[width=0.4\paperwidth]{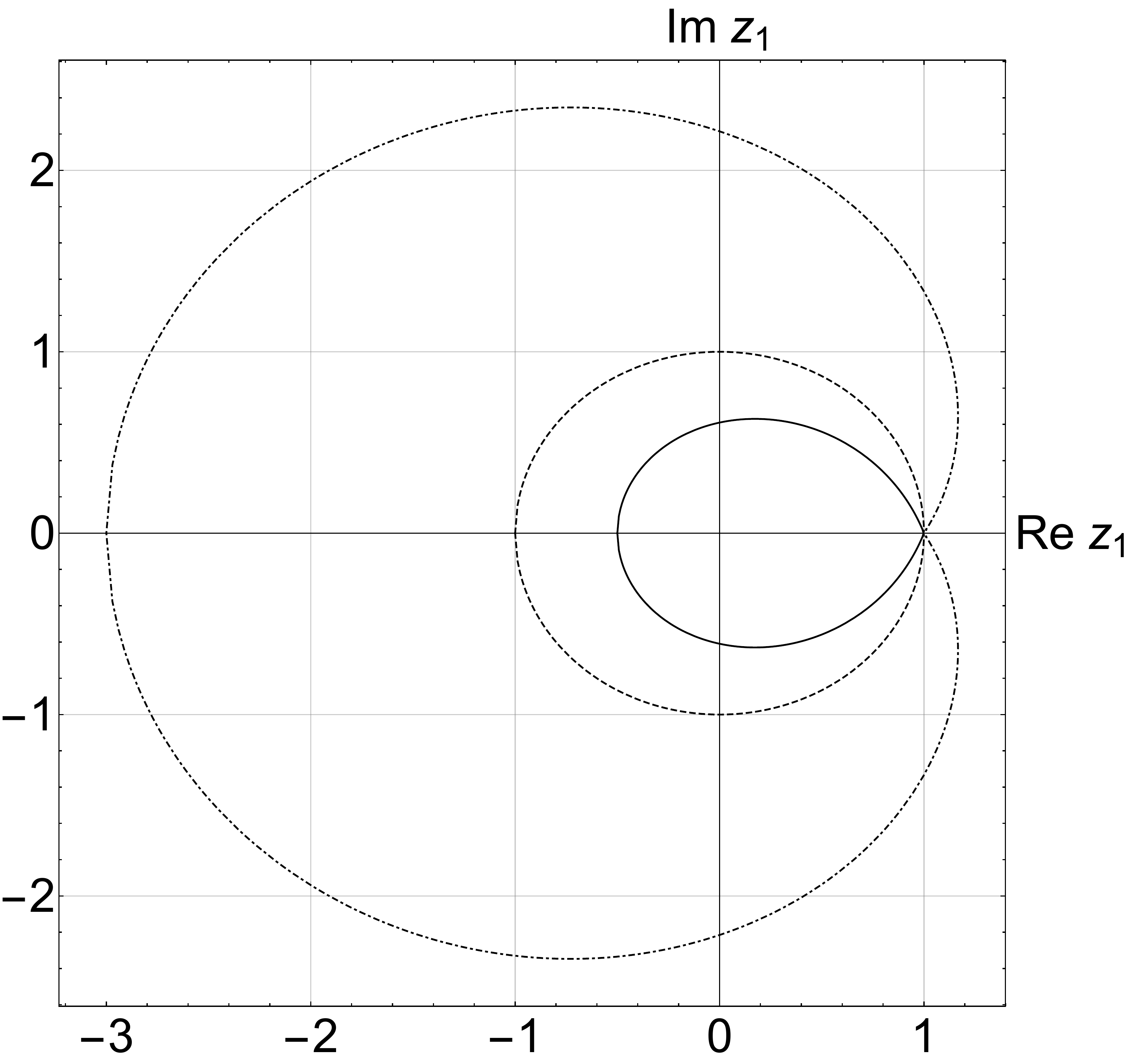}
	\caption{Edge loci in the complex $z_1-$plane for three Potts models ($q=1.5, q=2, q=4$, moving inside out) without invisible states ($r=0$).
	Each locus spans the full range of temperature values ($0\leq T\le \infty$) and intersects the ${\rm Re}\,z_1$ axis at $T=0$. 
		\label{r0yle}}
	\end{center}
\end{figure}

To more compactly illustrate the dependencies of zeros on the both temperature and on the number of states, instead of plotting the loci of the full sets of zeros as in Fig.~\ref{fig1}, we plot the coordinates of the edges for different values of $T$ and $q$ in Fig.~\ref{r0yle}. 
We call these ``edge loci''.
Such plots allow us to capture a greater span of $q$ and $T$ values while keeping the essential information because where the edge loci cross the real axis is where a phase transition can happen.
For different given values of $q$ the edge loci form different closed curves. 
But in each case the real axis is crossed at $T=0$ confirming that the only possibility is for a phase transition at zero temperature, as observed nearly a hundred years ago by Ising  (in the $q=2$ case) \cite{Ising}.


To summarise, in this subsection we have recovered known results, supporting the viability of the approach.

\subsection{Lee-Yang zeros for the Potts model with invisible states}
\label{III12}

Having recovered well-known results for the ordinary Ising model in Fig.~\ref{fig1} and the ordinary Potts model in Fig.~\ref{r0yle},
and illustrated how the loci depend on $q$,  we turn our attention to the Potts model with invisible states.
As follows from Eq.~(\ref{eigmain}) an external field acting on  invisible states effectively works as an additional number of such states. 
We elaborate on this duality in~\ref{Fisher} where we present a similar relationship between field and temperature.
Therefore, without loss of generality, we can set $h_2=0$ (or $z_2=e^{-\beta h_2}=1$) in Eq.~(\ref{eigmain}) arriving at 
\begin{equation}
\label{polyn.potts.inv}
(r-\lambda)(yz_1-\lambda-z_1)(y-\lambda-1)-\lambda z_1(y-\lambda-1)-(q-1)(yz_1-\lambda-z_1)\lambda=0.
\end{equation}
Setting $z_2=1$ in Eq.~(\ref{largeeq}), and using the method described earlier, 
we extract the Lee-Yang zeros in the complex $z_1$-plane for any value of $T$ at fixed values of $q$ and $r$. 
As a counterpart of Fig.~\ref{fig1} for the ordinary Potts model 
we plot Lee-Yang zeros of the $(2,2)$-state Potts model for different temperatures in  Fig.~\ref{q2r2}.

\begin{figure}[t]
	\begin{center}
		\includegraphics[width=0.4\paperwidth]{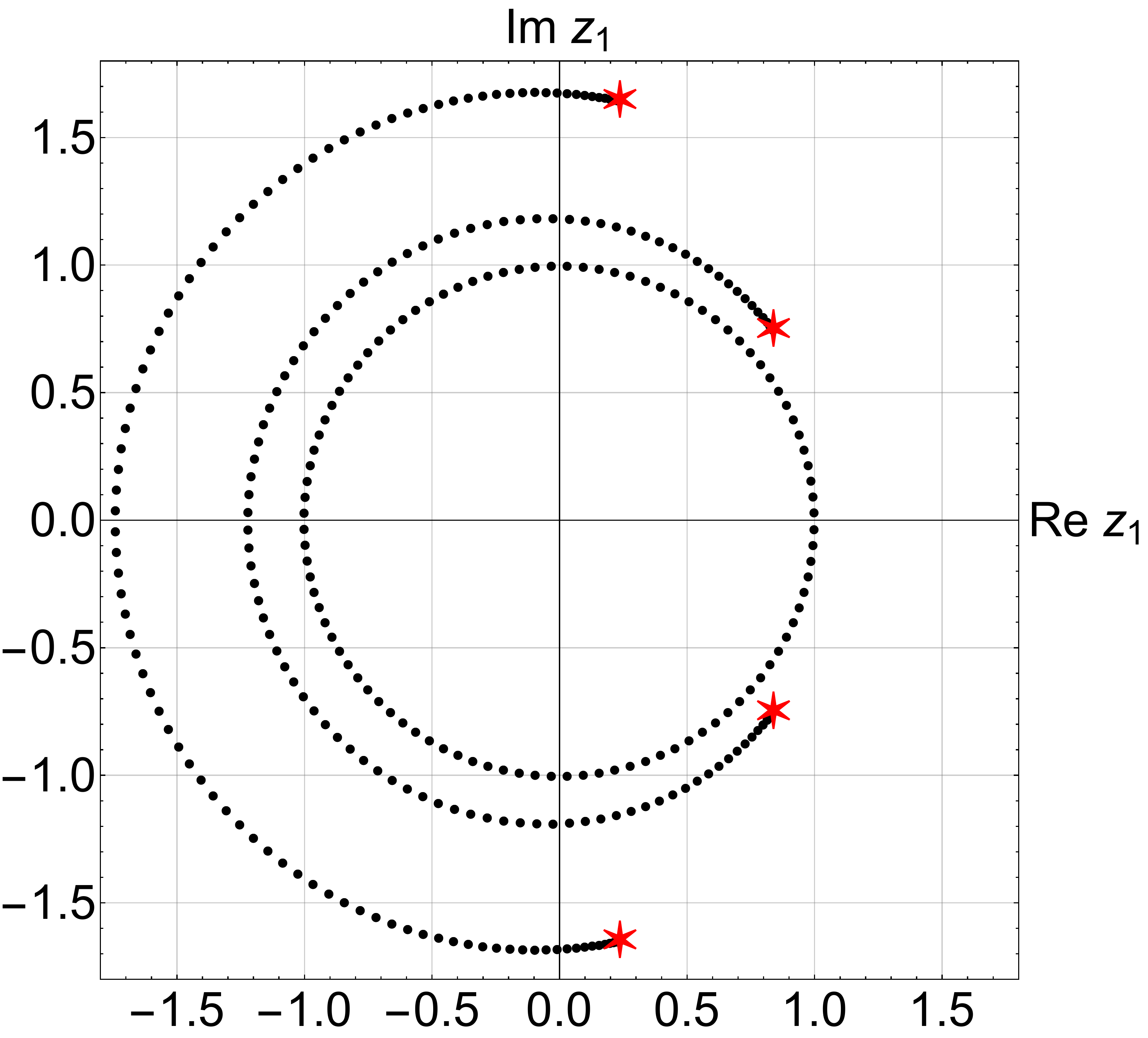}
		\caption{Lee-Yang zeros for the $(2,2)-$state Potts model in the complex $z_1=e^{-\beta h_1}$ plane at $y=e^\beta=2$ (outer  locus), $y=4$  and $y\rightarrow \infty$ (inner, closed circle, representing $T=0$) for systems of size $N=100$.
	The inner circle is identical to the right panel of Fig.~\ref{fig1} for the Ising model.  
	As in the left panel of Fig.~\ref{fig1}, the outer two loci indicate there is no transition at non-zero temperature.
	The Yang-Lee edges are highlighted in red.
			\label{q2r2}}
	\end{center}
\end{figure}

As seen from the plot, zeros form circular arcs, but their radii are not unity and increase with increasing temperature (lower $y$-values) as the system is driven further away from the (zero-temperature) phase transition. 
The same behaviour was observed even in the ordinary Potts model (Fig.~\ref{r0yle}). 
The difference is that even in the Ising case ($q=2$) the presence of the invisible states changes the radius of the circle. 
It is only at $T=0$ that zeros lay on the closed circle of unit radius.

It also is worth noting here that Lee-Yang zeros become increasingly dense as the edge is approached.
This behaviour is quantified by the edge singularity exponent $\sigma$~\cite{KoGr71}. 
The precise value of this exponent is discussed in \ref{Singular}.

To find the Yang-Lee edge one has to identify when the two largest eigenvalues of the transfer matrix are equal. 
This means that the polynomial (\ref{eigmain}) also has degenerate roots. 
This condition is equivalent to the discriminant $\mathbb{D}(y,z_1,z_2,q,r)$ of Eq.(\ref{eigmain}) vanishing:
\begin{equation} 
\label{discriminant}
\mathbb{D}(y,z_1,z_2,q,r)=0\,.
\end{equation} 
The discriminant $\mathbb{D}$ is a polynomial function of its arguments $y$, $z_1$, $z_2$, $q$ and $r$. 
By setting $z_2=1$ and fixing the numbers of visible  and invisible states ($q$ and $r$, respectively) we can scan values of the  temperature ($y=e^\beta$) to determine the coordinates of the edge points. 
Finding these coordinates for all possible temperatures $0\leq T < \infty$ (or $1\leq y<\infty$) we obtain the edge loci 
as shown in Fig.~\ref{q2ryle}.
This is the counterpart of Fig.~\ref{r0yle} (it is the Ising model with invisible states).
Comparison between the two figures illustrates that invisible states (entropy augmentation the energy lives in the density) are manifest in Lee-Yang-zero terms by widening the edge loci.
But the behaviour of Lee-Yang zeros discussed above signals that the presence of invisible states does not change the fact that there is only a zero-temperature  phase transition.

\begin{figure}[t]
	\begin{center}
    \includegraphics[width=0.4\paperwidth]{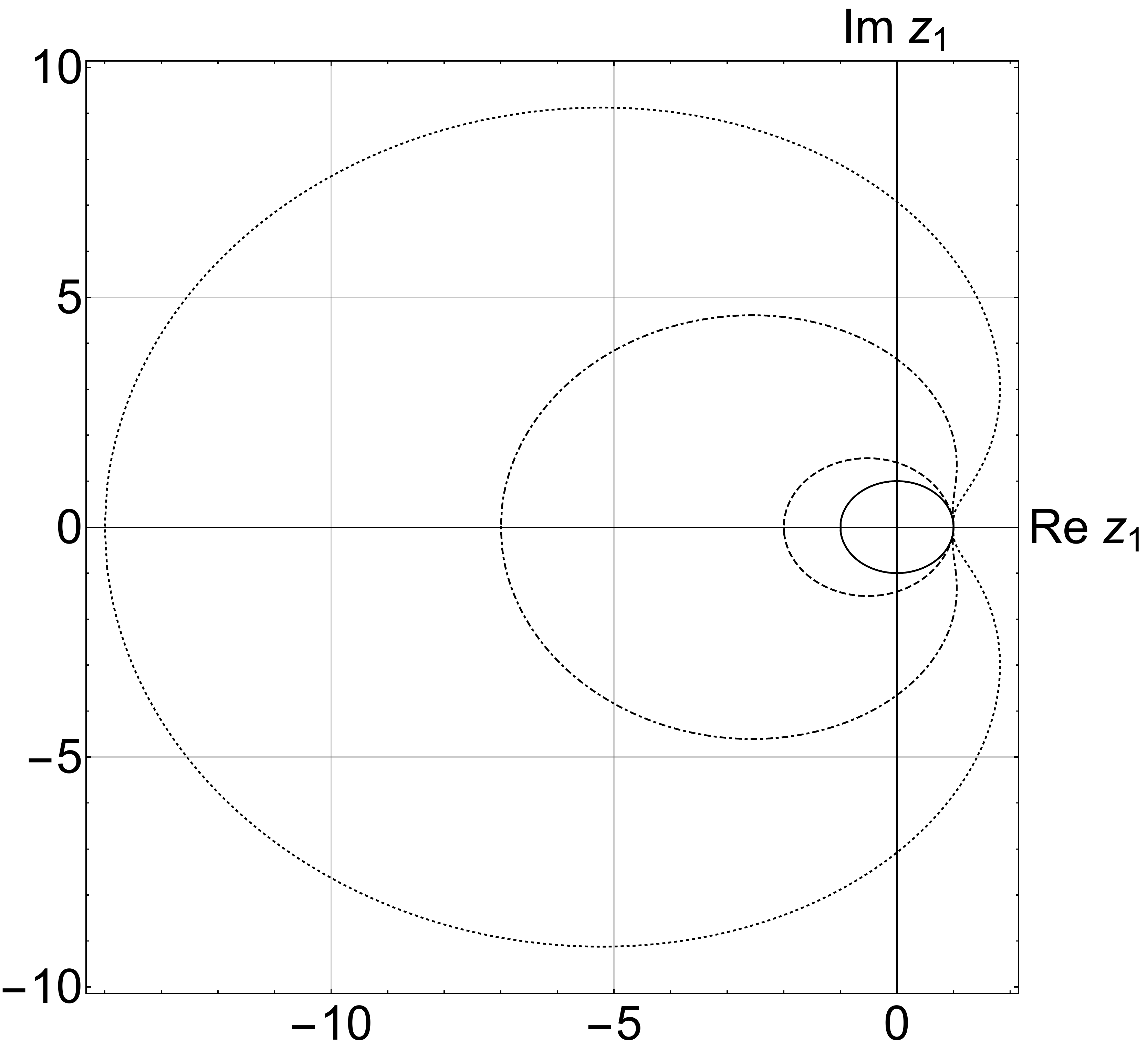}
	\caption{The edge loci in the complex $z_1$ plane for $q=2$ and $r=0$, $1$, $6$, $13$ moving inside out. 
	All plots cross at the real axis at $z_1=1$ indicating only zero temperature phase transitions.
	This is the invisible-states counterpart of Fig.~\ref{r0yle}.
	\label{q2ryle}}
	\end{center}
\end{figure}

To summarise, in this subsection we have shown that systems with a positive number of invisible states also fail to manifest a positive-temperature phase transition in 1D.

\section{Phase transitions at  positive temperatures}
\label{IV}

In this section we introduce new ways to instigate positive-temperature phase transitions in a one-dimensional classical model with short-range interactions. 

The original considerations of Landau and Lifshitz established that the free energy of a one-dimensional system with two mixed phases decreases with  increasing the number of  domain walls, so that an ordering phase transition is not possible for positive temperature \cite{Landau}.
An archetypal approach to circumvent this restriction is to increase the range of interaction and hence internal energy. A new way to achieve the same end is  to introduce a mechanism that leads to entropy decrease and therefore to a free energy increase.
The calculations displayed in this section are based on the obvious observation 
that if adding positive numbers of
invisible states increases the entropy, negative numbers of invisible states decreases it.

We introduce two ways to achieve positive-temperature transitions;
in Subsections~\ref{IV.0} and \ref{IV.1} we relax the condition of positivity on $r$
and
in Subsection~\ref{IV.2} we allow the external magnetic field $h_2$ to be complex.
Although both of these conditions are exotic, they either have connections to physical systems or have potential to be manifested physically in the future  \cite{Wei2012,Peng2015,Wang}. 


\subsection{Entropy depletion: Zeros in complex $z_2$ for $h_1=0$}
\label{IV.0}

To begin our investigations we take inspiration from the analysis of so-called Potts zeros. 
In the ordinary Potts model, these are studied by promoting the Potts variable $q$ to a complex number. 
Zeros in the complex $q-$plane are used to find the critical number of states for a given temperature \cite{Kim2001,Ghulghazaryan2003,Glumac2002}. 
While values of $q$ below $2$ are unphysical in terms of spin models, they can have 
physical manifestations --- for example $q=1$ describes percolation, $q=0$ spanning trees and Abelian sandpile model for self-organised criticality \cite{q1,q0}.

We have seen earlier that $r$ acts similarly to an external magnetic field $h_2$. 
One can therefore interpret zeros in the complex $r$ plane as Lee-Yang-type zeros in the complex plane of $z_2$.
We obtain partition function zeros in the  complex $z_2$ plane by substituting  values of $q$, $r$, $h_1$ into Eq.~(\ref{largeeq}) in a similar manner as in the previous section. 
On the other hand, we can analyse the behaviour of the edge coordinates directly by solving Eq.~(\ref{discriminant}). 
The corresponding plot is given in Fig.~\ref{z2zeros} for the particular case $q=2$.

\begin{figure}[t!]\begin{center}
\includegraphics[width=0.4\paperwidth]{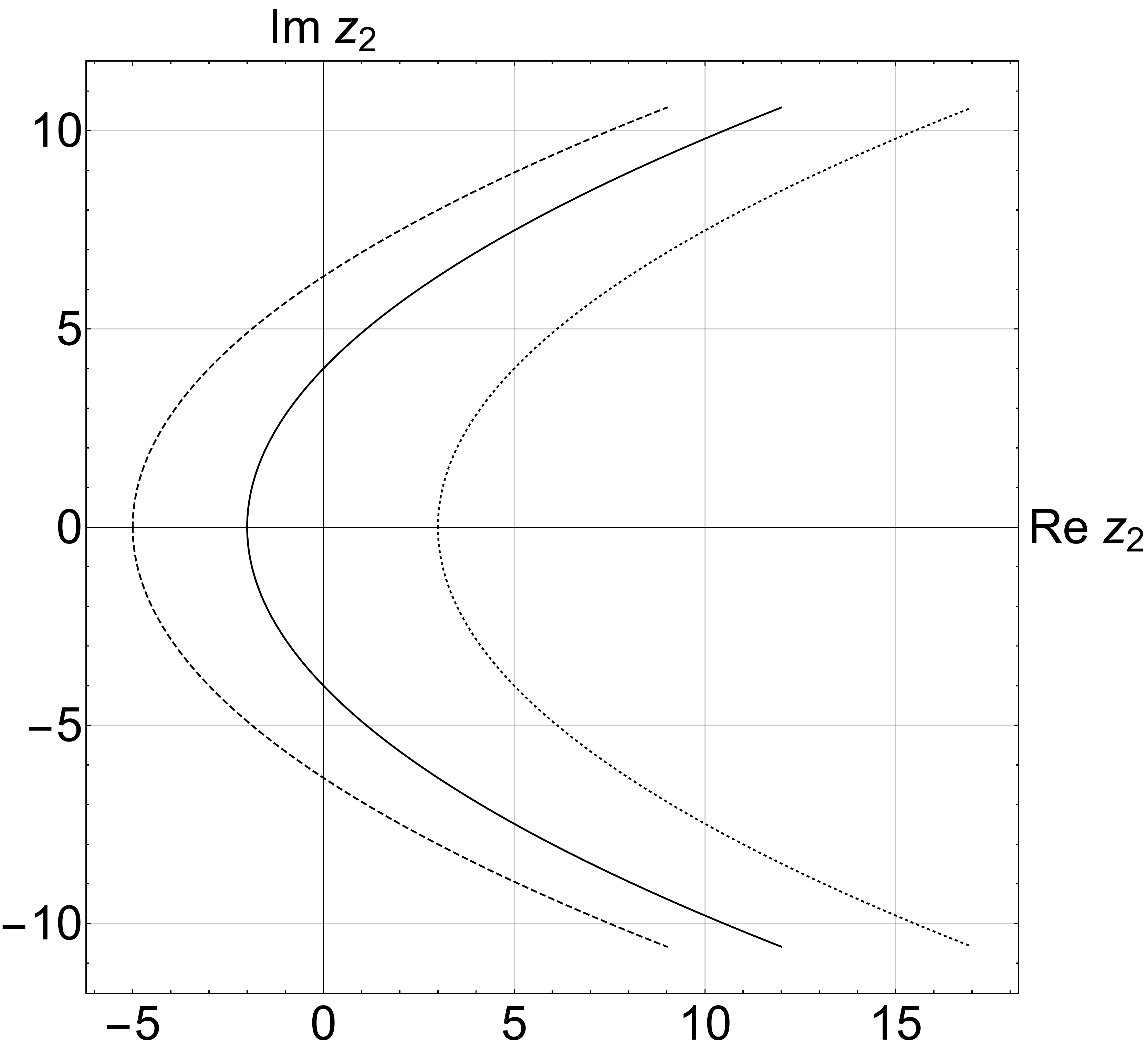}
	\caption{Yang-Lee edge locus corresponding to complex $z_2$ for $q=2$, $h_1=0$ and $r=4$ (dashed line), $r=1$ (solid line)  and $r=-4$ (dotted line). Each point of the lines represent a certain temperature in the region $0\leq t\leq 1$. 
	\label{z2zeros}}
\end{center}
\end{figure}

Spontaneous symmetry breaking (even at zero $T$) can only occur in the interacting system and, the fact that the field $h_2$ is conjugate to the non-interacting part of the Potts variable explains why  there is no phase transition in this subspace.
This is illustrated in Fig.~\ref{z2zeros} where there is no intersection point at the positive real semi-axis for positive values of $r$. 
The middle edge locus in Fig.~\ref{z2zeros} was obtained for $r=1$. 
Increasing $r$ serves to shift the locus to the left; but its shape remains the same. 
This confirms our interpretation of Fig.~\ref{q2ryle} that increasing the value of $r$ pushes the loci further away from a physical phase transition.
One may infer that the converse also holds; decreasing $r$ to a negative number of invisible states may shift the real edge locus from the negative real semi-axis to its positive counterpart. 
This is supported in the rightmost locus of Fig. \ref{z2zeros}.

To summarise this section, we have observed that zeros in complex $h_2$ can cross the real axis  provided $r$ is negative.
Of course, the figure refers to the complex $z_2$ plane, corresponding to a field acting on entropic (invisible) states only.
To connect with previous studies of Lee-Yang zeros we have to examine the complex $h_1$ or $z_1$ plane. We start in Subsection \ref{IV.1} by setting $h_2=0$ and defer complex $h_2$ to Subsection \ref{IV.2}.

\subsection{Entropy depletion: Zeros in the complex $z_1$ plane for $h_2=0$ and $r<0$}
\label{IV.1}

We build upon the observation in Fig.~\ref{z2zeros} that the edges in the complex $z_2$ plane are horizontally shifted to the right by reducing $r$ to negative values. 
Therefore  we relax the condition of positivity on the number of invisible states $r$
when, in Fig.~\ref{LYminusr}, we plot the Lee-Yang zeros in complex $z_1$ for the $(2,-5)-$state Potts model for different values of temperature.  
Fig.~\ref{LYminusr} is a negative-$r$ counterpart of Figs. \ref{fig1} and \ref{q2r2}. 
Our principal result in this subsection is that loci of zeros cross the real axis at positive temperatures for negative values of $r$. In particular, the left plot in the second line shows that, in the absence of field $h_1$ (i.e., when $z_1=1$), the zeros cross the real axis at a positive value of $t$ (namely $t=0.25$).  
This is the sought-after spontaneous, zero-field, positive-temperature phase transition in 1D. 
The correlation length is infinite at the critical temperature but the entropy has discontinuity. 
Therefore the phase transition can be interpret as of first order. However, small part of the entropy dependency on the temperature has an unphysical region. This can be as a consequence of unphysical values of the model parameters.

\begin{figure}[t!]
	\begin{center}
		\includegraphics[width=0.3\paperwidth]{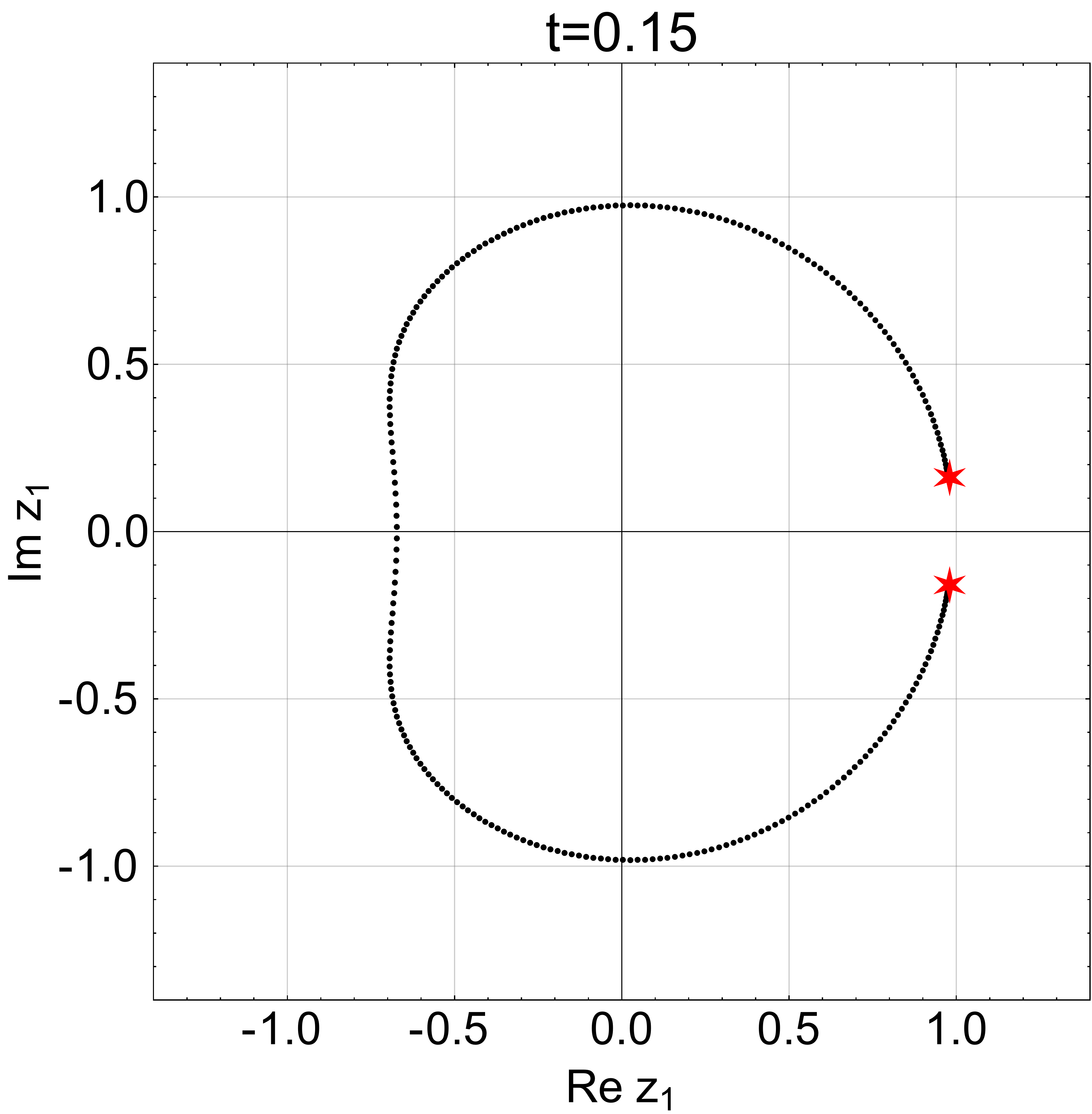}
		\includegraphics[width=0.3\paperwidth]{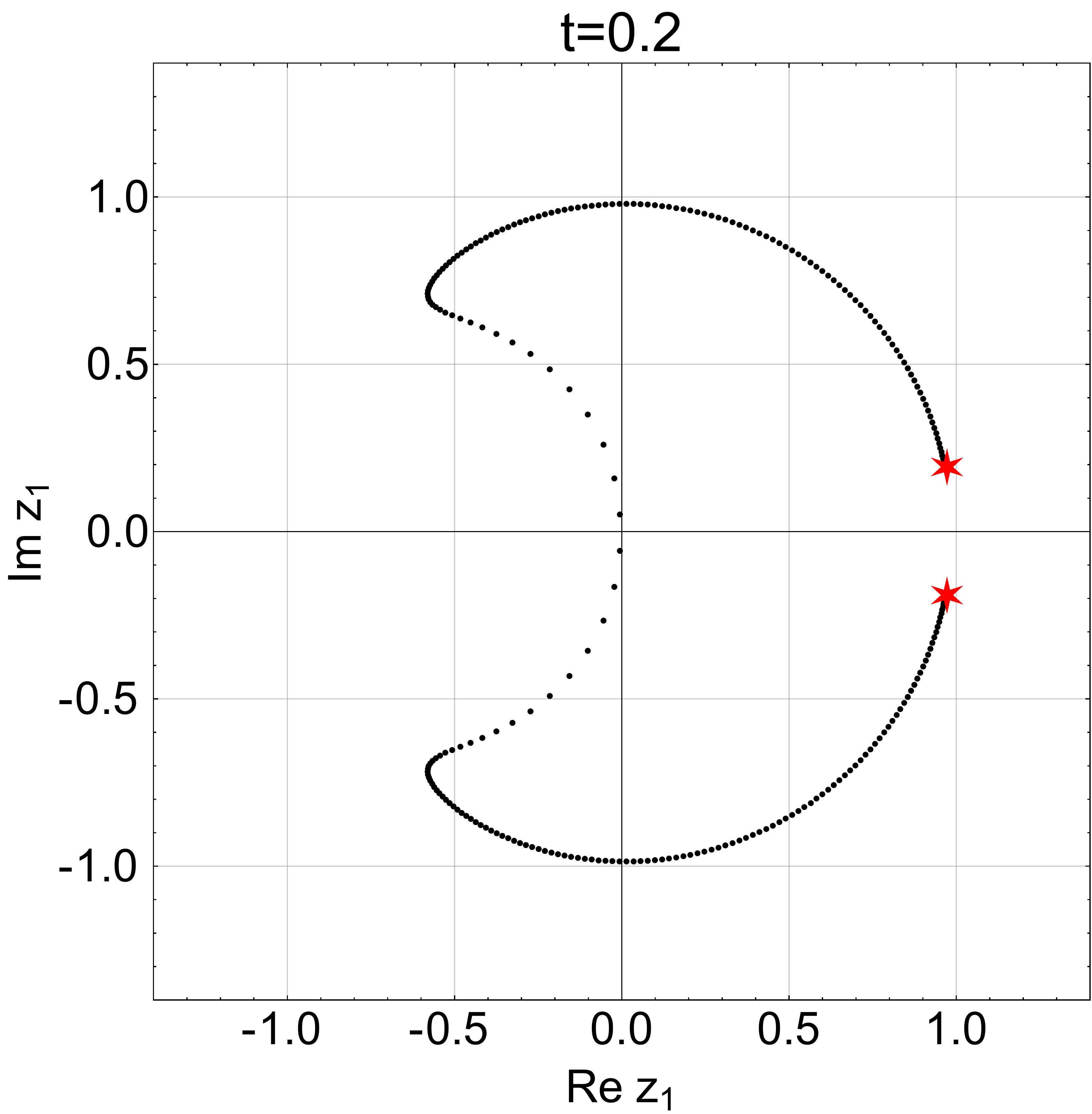}\\
		\hspace*{0.04\textwidth}(a)\hspace*{0.39\textwidth}(b)\\
		\vspace*{2mm}
		\includegraphics[width=0.3\paperwidth]{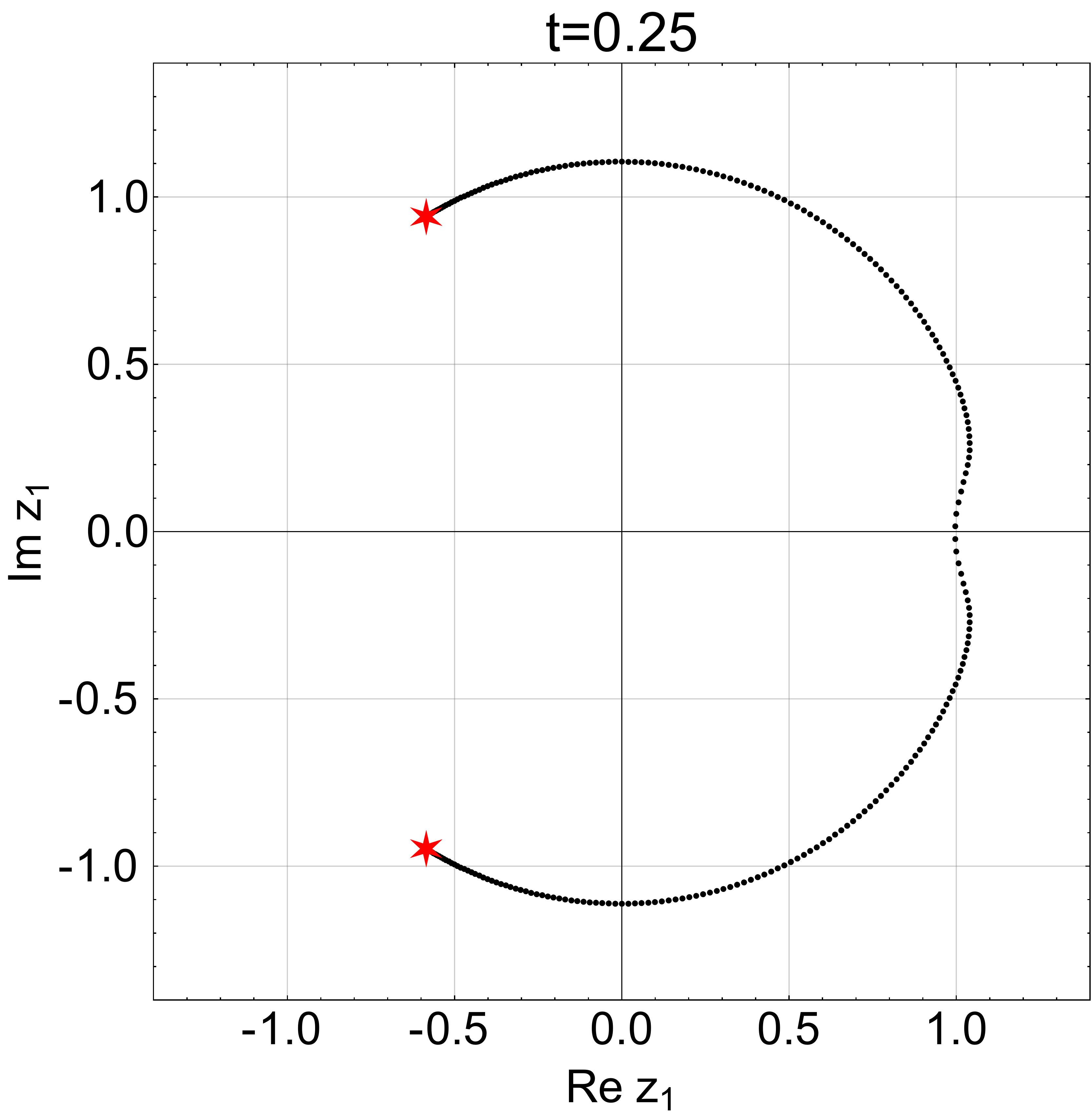}
		\includegraphics[width=0.3\paperwidth]{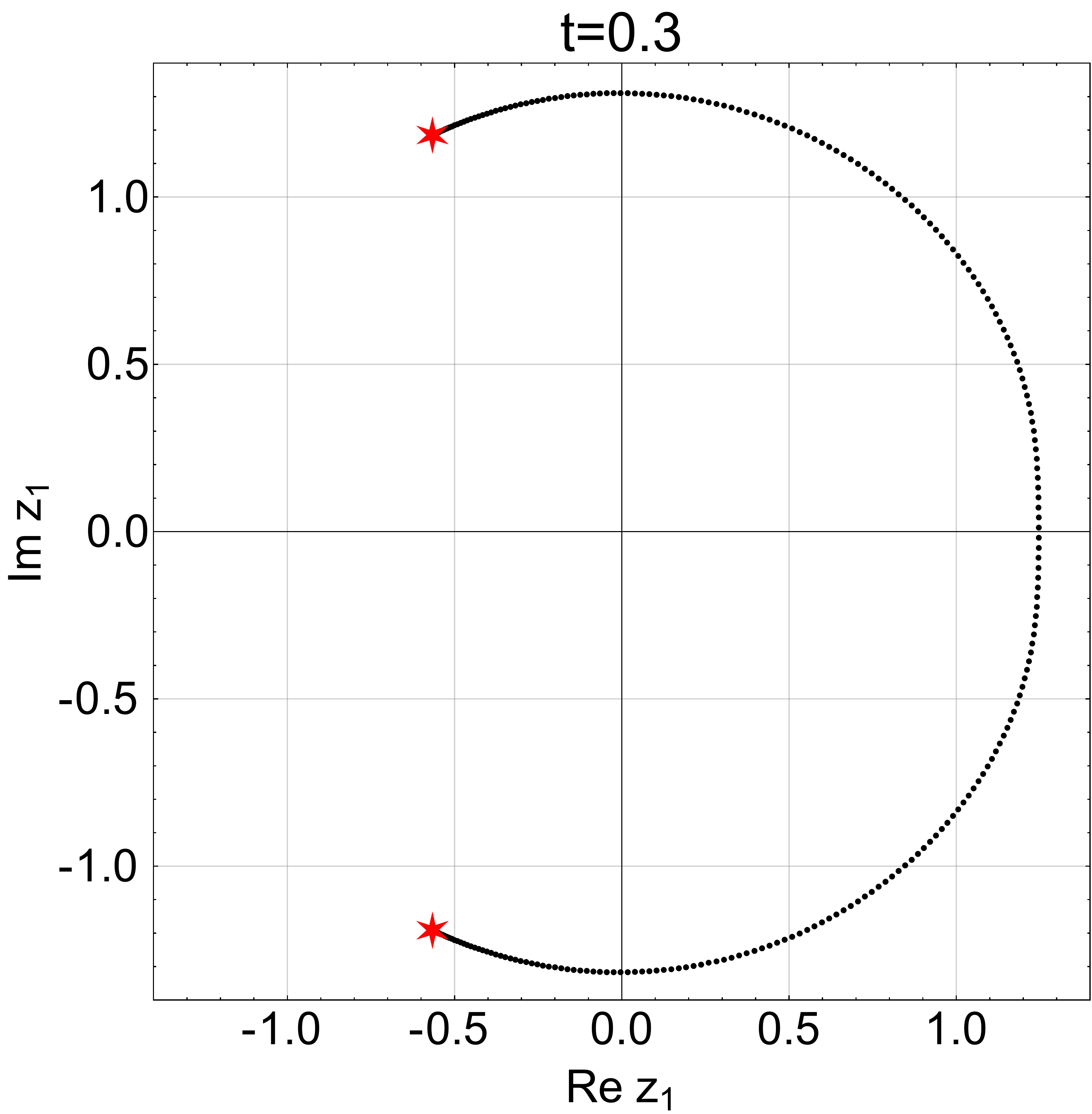}\\
		\hspace*{0.04\textwidth}(c)\hspace*{0.39\textwidth}(d)
		\caption{Lee-Yang zeros of the $(2,-5)$-state Potts model at different values of temperature (a)~$t=0.15$, (b)~$t=0.2$, (c)~$t=0.25$, (d)~$t=0.3$ in the complex $z_1-$plane for the system size $N=256$. Panels (a) and (b) illustrate zeros below the zero-field critical temperature $t_c=0.25$, panel (c) and (d) illustrate zeros at and above $t_c$. Large red stars show edges and black dots ordinary Lee-Yang zeros. For small temperatures edge is located in the positive semi plane $\rm{Re}\,z_1>0$, while at higher temperature it jumps to the region $\rm{Re}\,z_1<0$. This jump occurs below the critical temperature $t_c$. 
		\label{LYminusr}}
	\end{center}
\end{figure}

Each of the loci in Fig. \ref{LYminusr} crosses the real axis, corresponding to  a critical point.
Negative values of $z_1$ represent complex values of physical field $h_1$.
Values $z_1>1$ represent $h_1>0$ --- a positive field acting on on the first state $s=1$.
Positive values in the range $0<z_1<1$ corresponds to negative values of $h_1$.
Such negative values of the external magnetic field are effectively the same as positive external fields acting on the other ($s \ne 1$) states. 
For the ordinary Ising case ($q=2, r=0$) negative values of the external magnetic field acting on the first state ($s=1$), say, effectively represent the same physics as a positive external field acting on the other ($s=2$) state.
For the ordinary Potts model with $q>2, r=0$, negative external fields disfavour one of the states reducing the symmetry from $Z_q$ to $Z_{q-1}$. In three dimensional three state Potts model this affects the phase diagram; 
weak magnetic fields do not change the order of the phase transition, while strong negative magnetic field change it to the three-dimensional (3D) Ising universality class \cite{Bonati2010}. 

The figure also illustrates that, as for the ordinary Potts model, the Lee-Yang circle theorem is violated for the Ising model with a negative number of invisible states; 
the loci of zeros are not circular.

The set of crossing points for various temperatures in the range $0\leq t \leq 1$ can be interpreted as a phase diagram and is shown for the $(2,-5)$-state model as {red{the}} solid black line in Fig. \ref{phasediagram}. 
The spontaneous transition is identified at $t=0.25$, $z_1=1$.
The counterpart for the ordinary Ising model is at  $t=0$, $z_1=1$ --- i.e., at vanishing instead of positive temperature. 
To further illustrate this representation, in Fig. \ref{phasediagram} we divide the $(t,z_1)-$plane into regions. 
The different colours represent different eigenvalues which are maximal by absolute values.
Where they coincide is where criticality occurs.

\begin{figure}[t]
	\begin{center}
		\includegraphics[width=0.4\paperwidth]{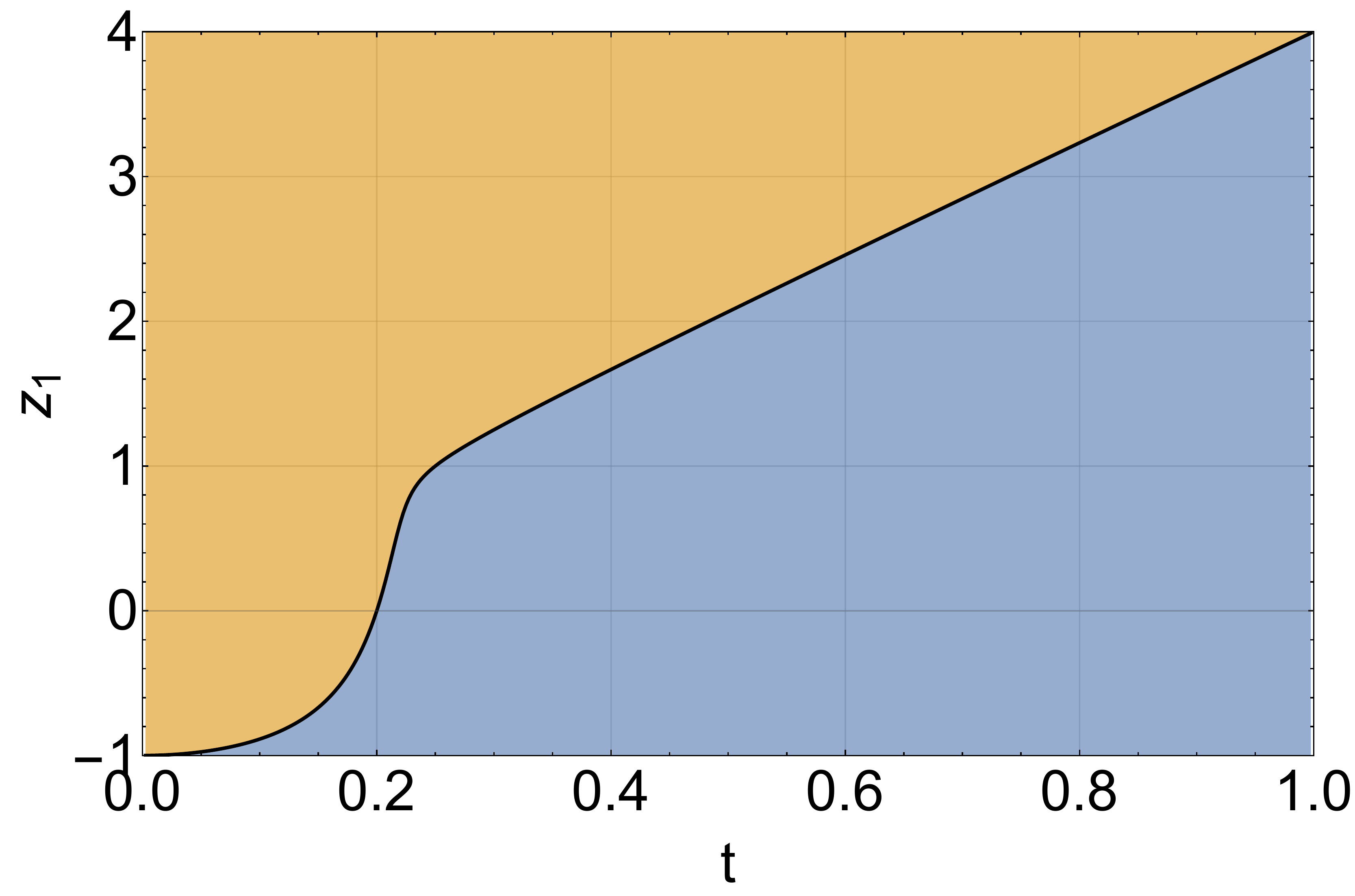}
		\caption{Phase diagram of the $(2,-5)$-state Potts model. Right panel: $(t,z_1)-$plane divided into regions according to the maximal eigenvalue. Values $z_1<0$ correspond to the complex values of magnetic field $h_1$, while $0<z_1<1$ corresponds to negative values of physical field $h_1$. 
			\label{phasediagram}}
	\end{center}
\end{figure}

A curious phenomenon in Fig. \ref{LYminusr} is the flipping with increasing $T$ of the edges of the loci (illustrated as large stars, red online)  from the positive to negative half planes.
This happens not at the zero-field critical point but at a lower value of $T$.
The reason for this is that the edges in Fig. \ref{LYminusr} are each away from the real axis and are pseudocritical points with ${\rm{Im}}\,z_1 \ne 0$ --- not zero-temperature critical points.
The phase diagram of Fig. \ref{phasediagram} has vanishing imaginary field  ${\rm{Im}}\,z_1 = 0$. 
To access the LY edges, and their flipping, requires non-zero values of ${\rm{Im}}z_1$ and three examples of this are depicted in Fig. \ref{phasediagram}.
Different colours in the plot represent different eigenvalues.
These are basically three slices through of a 3D plot with axes $t$ (temperature), ${\rm{Re}}\,z_1$
and ${\rm{Im}}\,z_1$. These plots are given for the fixed temperature, when flipping occurs and coordinates of the edge.
Three eigenvalues are equal by modulus exactly at the point where flipping occurs. 
Such a behaviour signals existence of a point with unusual Lee-Yang edge singularity exponent \cite{Dalmazi2008}.

\begin{figure}[t]
	\begin{center}
		\includegraphics[width=0.3\paperwidth]{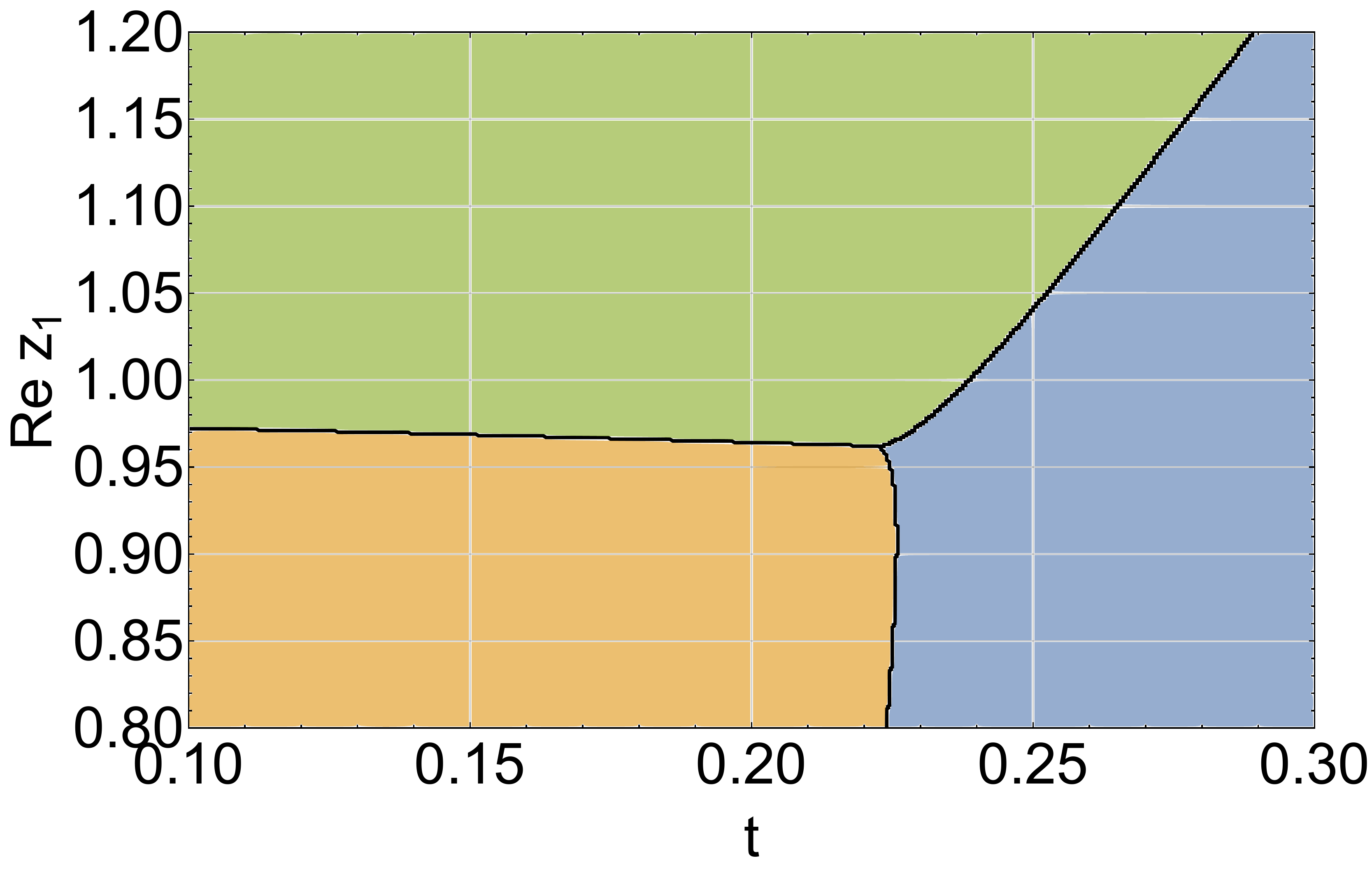}
		\includegraphics[width=0.3\paperwidth]{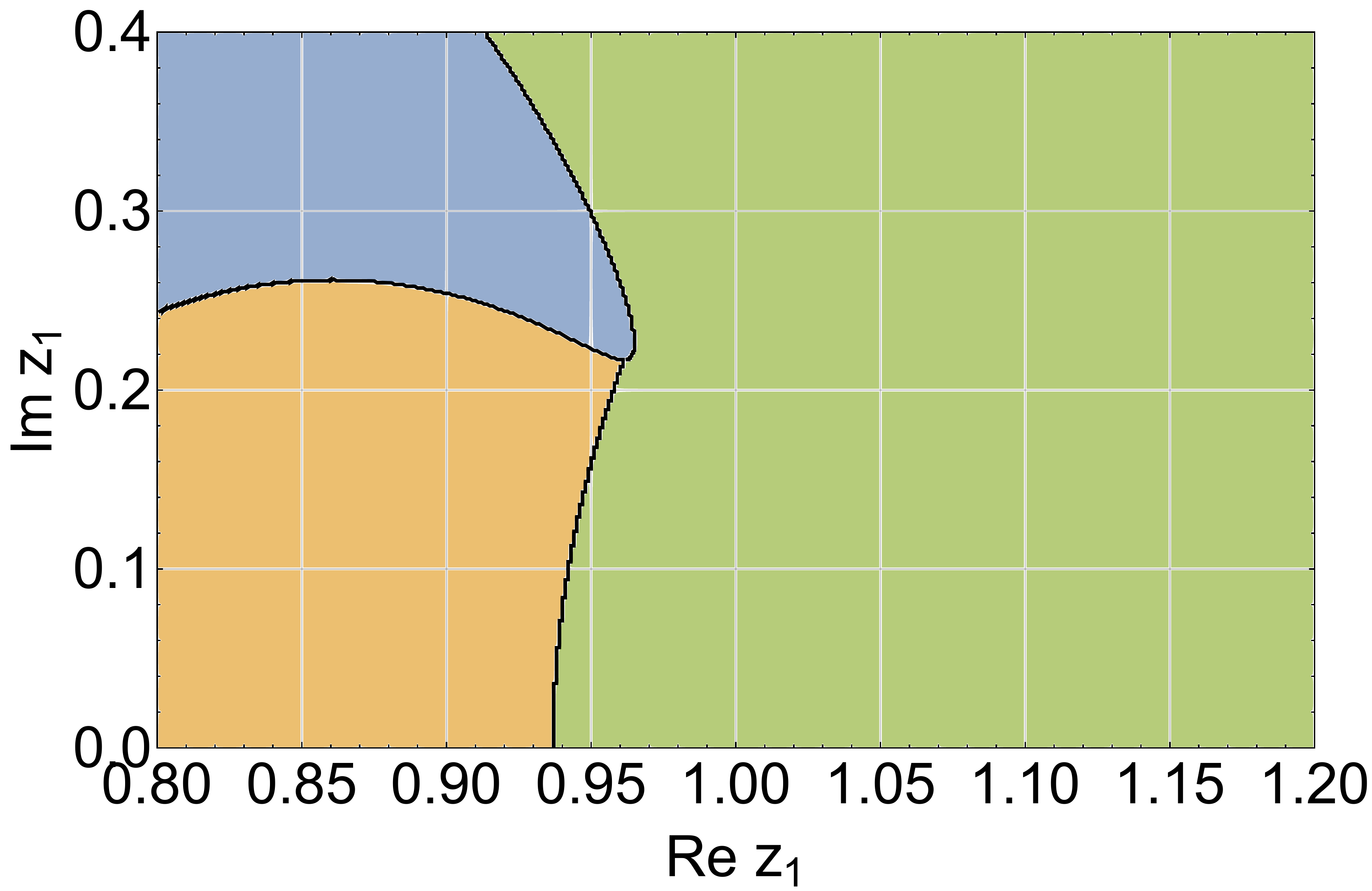}\\
		\quad(a)\hspace*{0.28\paperwidth}(b)\\
		\includegraphics[width=0.3\paperwidth]{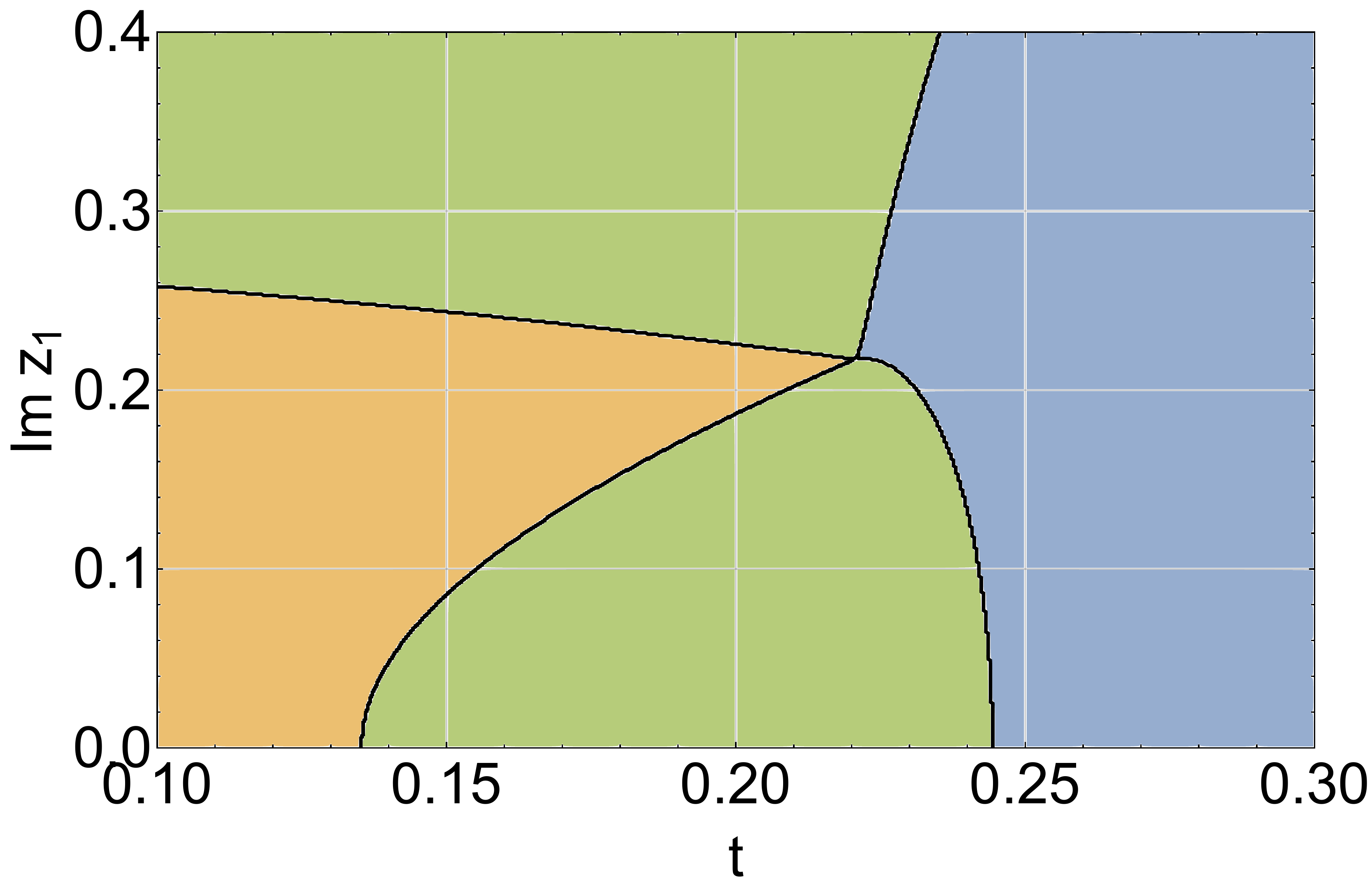}\\
		\quad(c)
		\caption{Three cross-sections of the 3D phase diagram. Each section is given at the fixed values of parameters when flipping of the edge occurs: (a) - fixed $\rm{Im}\,z_1$, (b) - fixed $t$, (c) - fixed $\rm{Re}\,z_1$. Colour of the region represent the eigenvalue, which is the largest by modulus inside this region. When three colours meet is the point where flipping of the edge occurs. 
			\label{curious}}
	\end{center}
\end{figure}



To summarise this section, while positive values of $r$ do not change the order and temperature of the phase transition, negative numbers of invisible states shift it to positive temperatures. 
In addition, there is a curious phenomenon involving the flipping of the locations of the edges relative to the other Lee-Yang zeros. This occurs at a value of temperature below the critical one and is explained by complex fields. We investigate complex fields in the next subsection.

\subsection{Lee-Yang zeros in the complex $z_1$ plane with a complex field acting on invisible states: The case $z_2<0$ ($h_2 \in \mathbb{C}$).}
\label{IV.2}

Following on from the above considerations, we extend our search for positive-temperature phase transitions in one dimension to an analysis of the effects of negative values of $z_2$ (meaning a complex external magnetic field $h_2$) through Lee-Yang zeros in the complex $h_1$ plane. 
Using the same method as previously deployed, we substitute into Eq.~(\ref{largeeq}) negative value of $z_2$  and obtain Lee-Yang zeros for various values of $q$, $r$ and temperatures $t$.  
Results are represented in Fig.~\ref{LYminusz2}, which is the positive-$r$ and negative-$z_2$ counterpart of Fig.~\ref{fig1} (which has vanishing values of $r$ and $h_2$). 
At small temperatures these zeros lay close to the unit circle of Fig.~\ref{fig1}, but increasing the temperature does not leave the Lee-Yang zeros on the unit circle.  
Instead they assume rather moon-like shapes. 
Moreover, although  the locus of Lee-Yang zeros opens with increasing temperatures as in Fig.~\ref{fig1}, the orientation of the arcs is reversed. 
It is worth noting here, that in order to obtain plots shown in Fig.~\ref{LYminusz2} we fixed $z_2$ and not $h_2$, meaning that with the change of temperature $t$ the external magnetic field is changed.

\begin{figure}[t]
	\begin{center}
		\includegraphics[width=0.3\paperwidth]{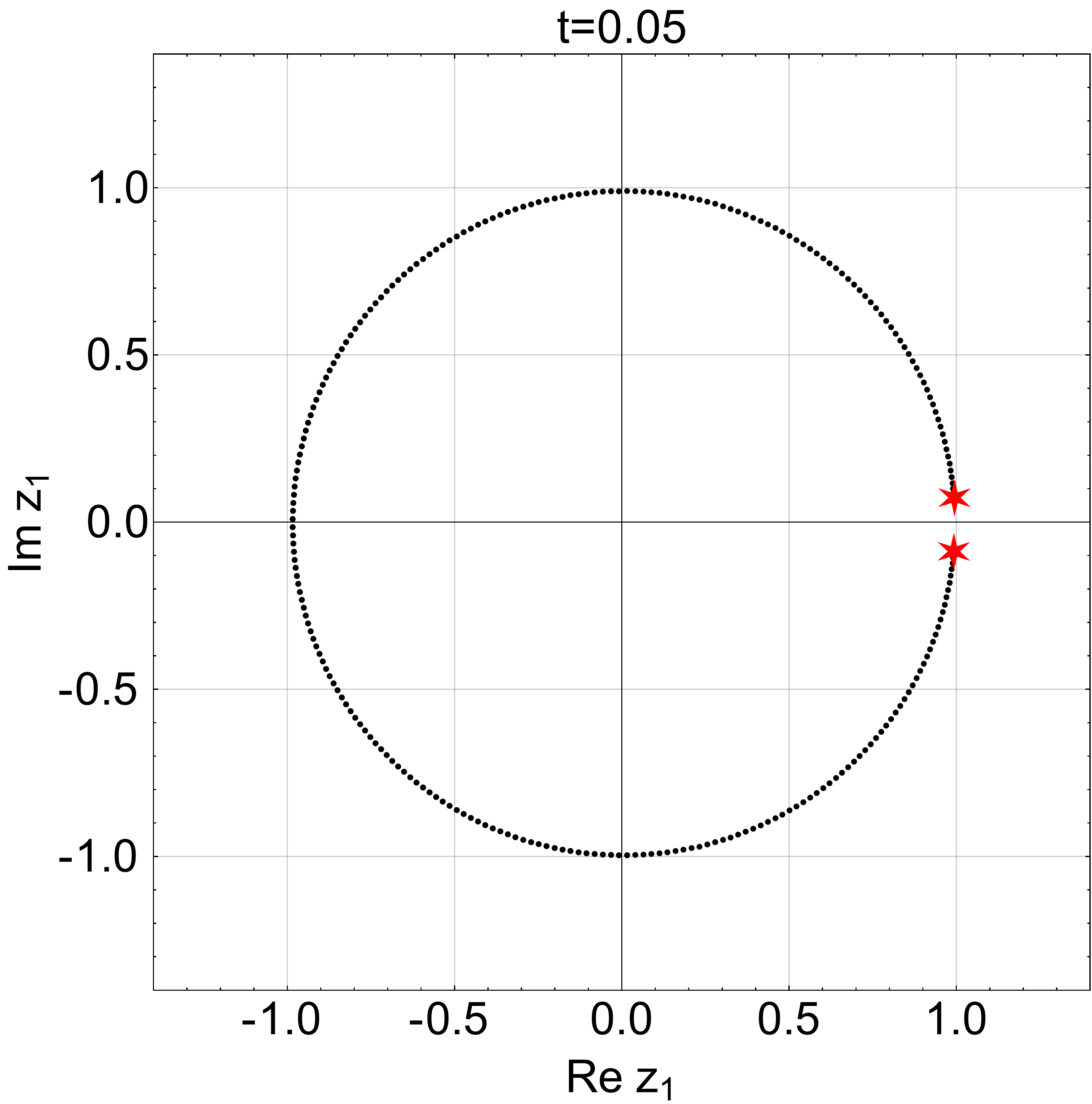}
		\includegraphics[width=0.3\paperwidth]{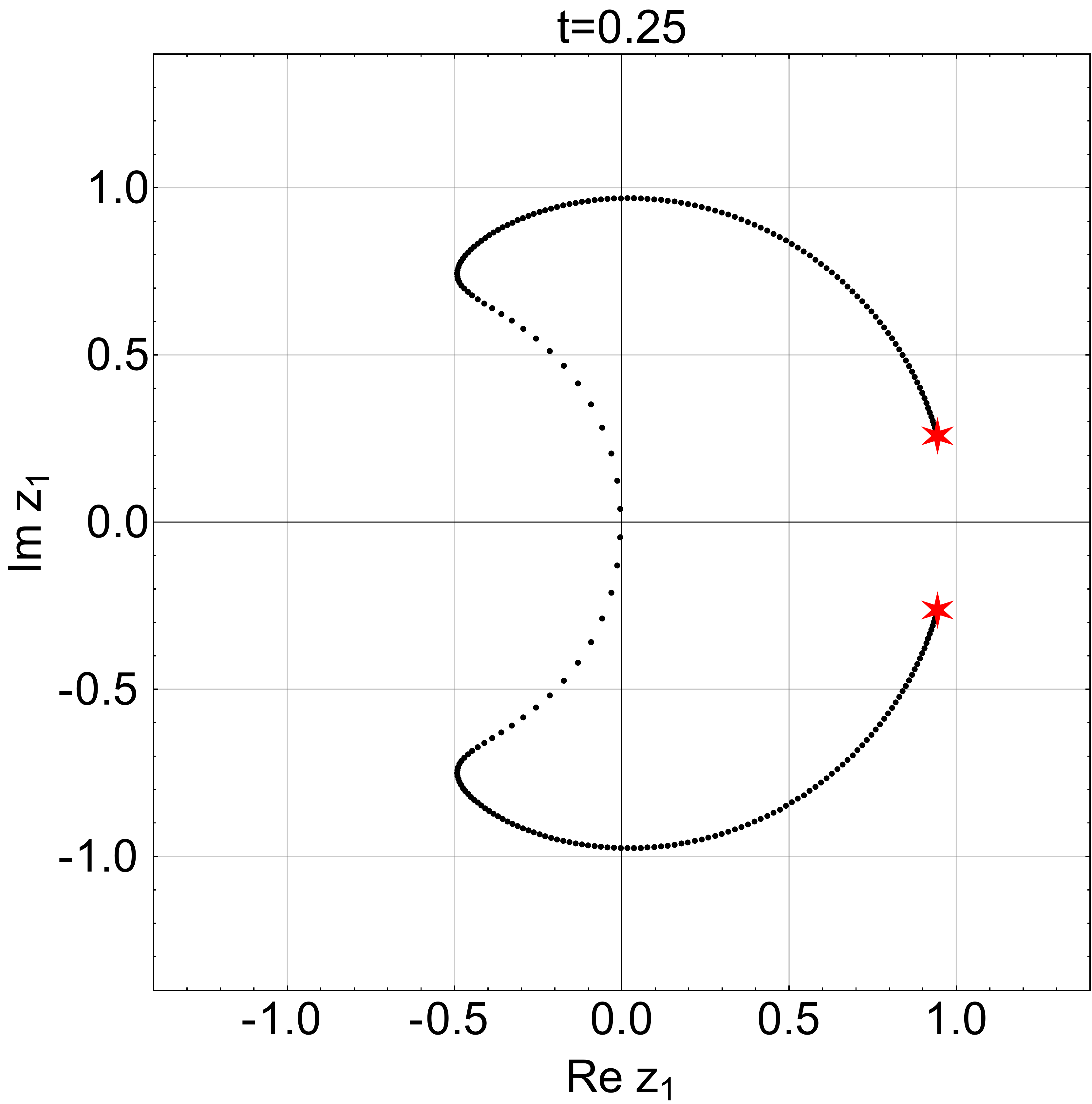}\\
		\hspace*{0.04\textwidth}(a)\hspace*{0.39\textwidth}(b)\\
		\vspace*{2mm}
		\includegraphics[width=0.3\paperwidth]{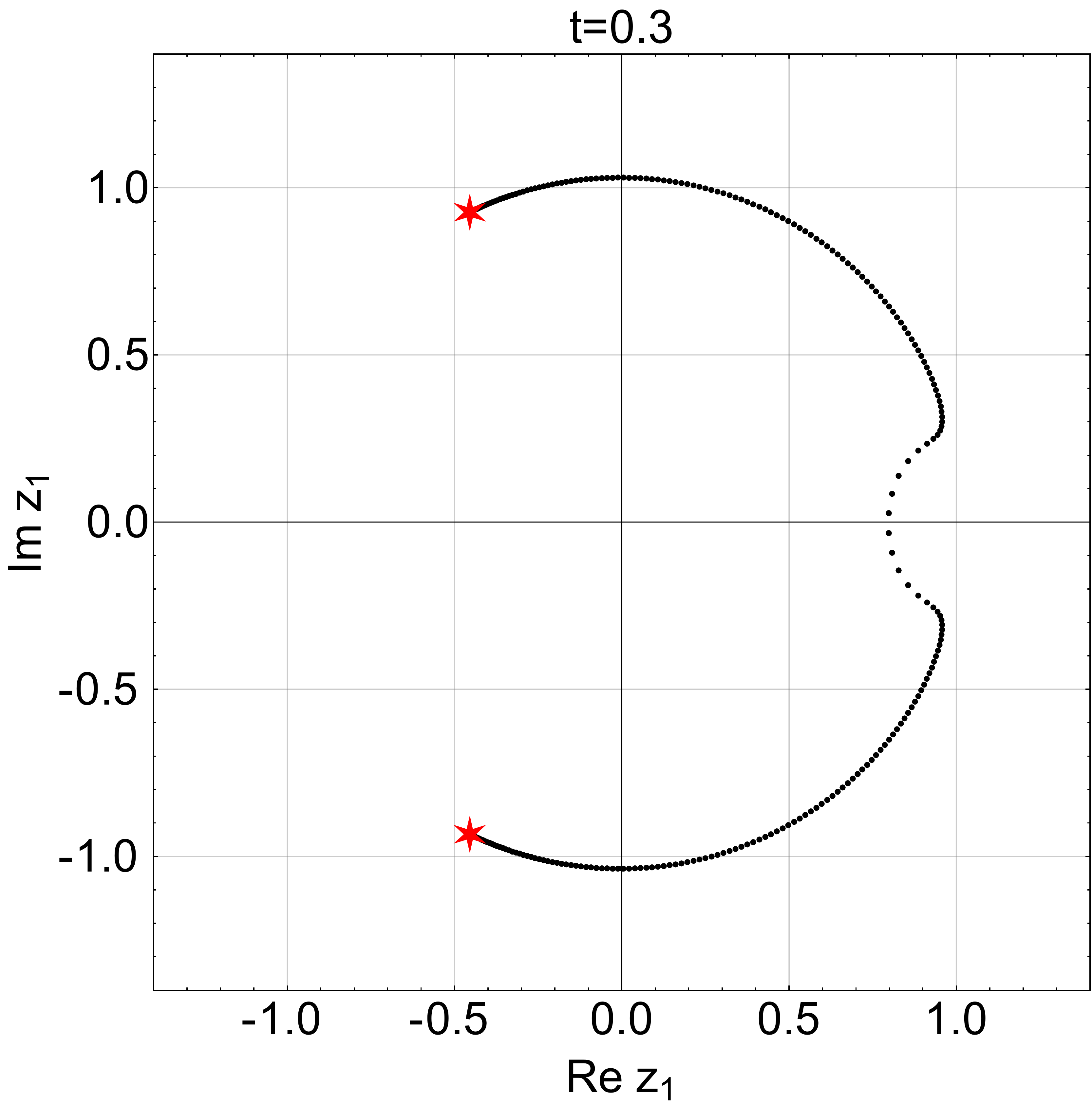}
		\includegraphics[width=0.3\paperwidth]{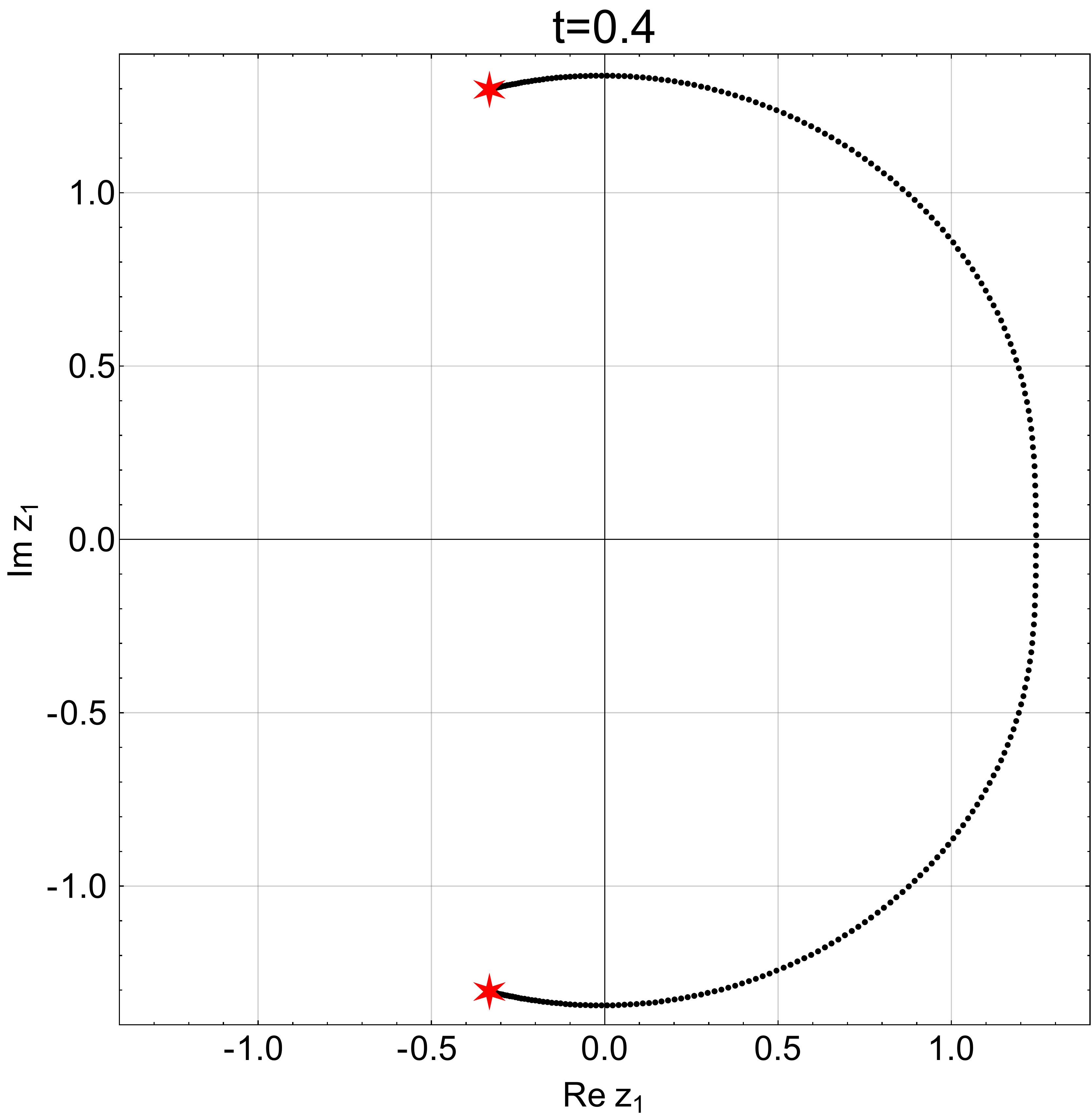}\\
		\hspace*{0.04\textwidth}(c)\hspace*{0.39\textwidth}(d)
		\caption{Lee-Yang zeros in the complex $z_1-$plane for the  $(2,2)$-state Potts model with $z_2=-5$ for different temperatures (a) $t=0.05$, (b) $t=0.25$, (c) $t=0.3$ and (d) $t=0.4$. Large red stars correspond to the edges. Overall behaviour of zeros is similar to that described in the previous subsection.
		\label{LYminusz2}}
	\end{center}
\end{figure}

To further analyse the complex values of $h_2$ that shift the phase transition to positive temperatures, we set $h_1=0$ and use the condition (\ref{discriminant}) for the discriminant to find when two eigenvalues are equal and largest by modulus.
This leads to a relation between the critical temperature and field $h_2$, namely
\begin{equation}\label{z2y}
z_2=y-q-r \pm 2 i \sqrt{q (y-1)}\, .
\end{equation}
In this equation both $y$ and $z_2=y^{h_2}$ are temperature dependent.
Fixing $q$, $r$ and sweeping through the region $1\leq y<\infty$ (meanting a temperature range $0\leq t \leq 1$), 
we solve numerically Eq.~(\ref{z2y}) and obtain complex values for $h_2$.
In Fig.~\ref{complexh2} we plot these values for the $(2,3)-$state Potts model in the form $e^{-\beta h_2}$. 
The curve forms two lines, each point of which corresponds to a certain positive critical temperature. The upper and lower branches correspond to complex conjugate values of the field.

\begin{figure}
	\begin{center}
		\includegraphics[width=0.4\paperwidth]{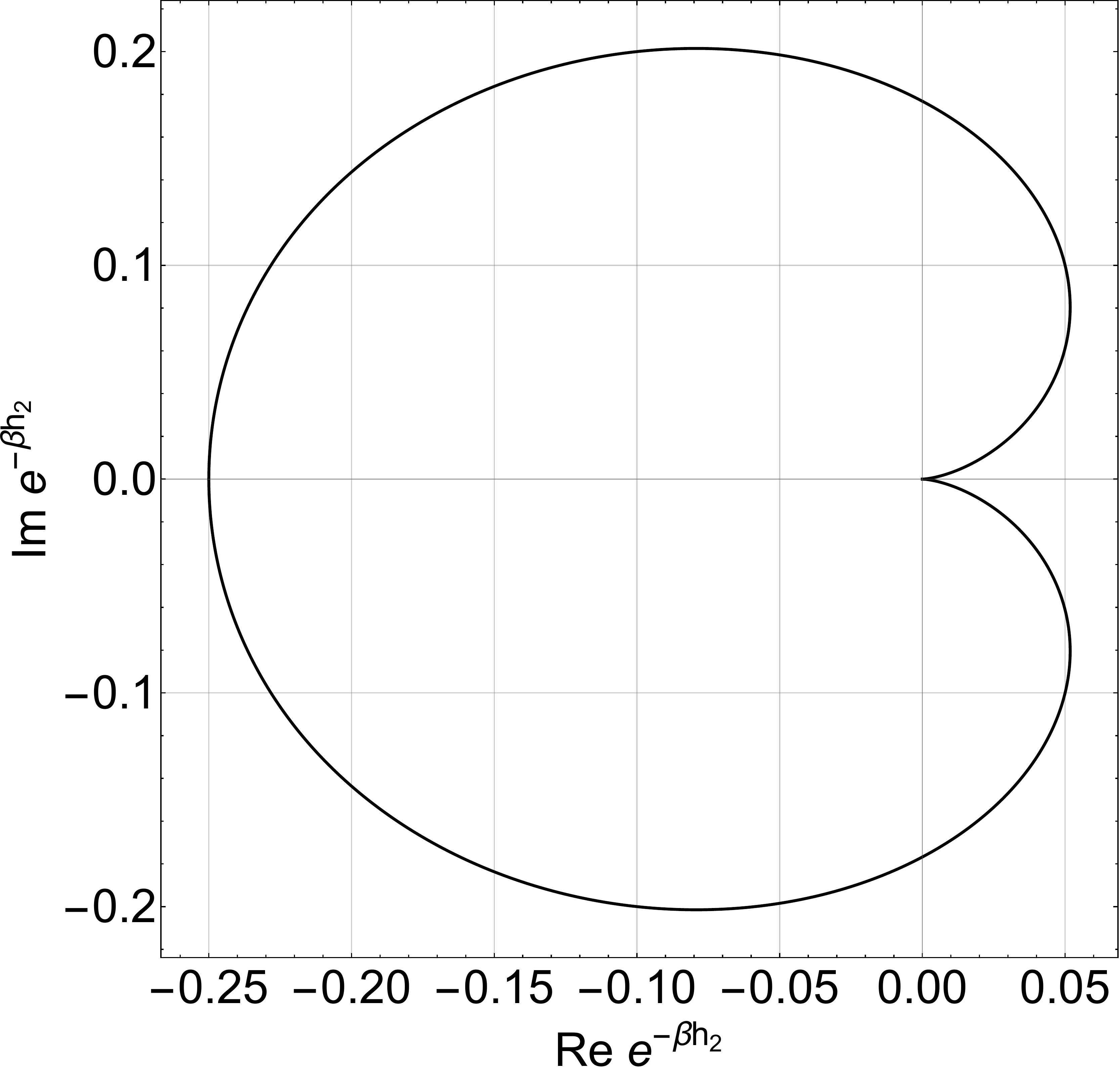}
		\caption{Values of $e^{-\beta h_2}$, for which the phase transition in the (2,3)-state Potts model occurs at positive
			temperature. Each point of the plot corresponds to a certain physically accessible critical temperature.}\label{complexh2}
	\end{center}
\end{figure}

To summarise this subsection, we have observed that a complex field acting purely on invisible states can induce a positive-temperature phase transition in 1D.
Although this appears exotic, 60 years after their introduction \cite{LY}, it has recently been established that complex external magnetic fields $h_1$ can be mapped into physically accessible quantum coherence times \cite{Wei2012}. 
Similarly, tuning complex values of the external field $h_2$ may one day be accessible, perhaps by changing the ``invisible'' part of a system's behaviour from classical to quantum.

\section{Conclusions}
\label{V}

The possibility of classical, equilibrium phase transitions below the lower critical dimension is important at a fundamental level as well as for potential  manifestations in real-world systems.
Onsager's solution of the 2D Ising model \cite{Onsager} was only recently confirmed experimentally \cite{LeLe16} and
theoretical investigations  have shown that adding invisible states can alter the type of phase transition present in such models \cite{Tamura2010,Tanaka2011,Krasnytska2016}. 
Other recent theoretical and experimental developments include the establishment of a link between complex fields and  quantum coherence times \cite{Wei2012,Peng2015}, opening up new ways to access complex fields physically \cite{Press}.
Here we combine all of these recent developments with an exact solution of the Potts model with invisible states on a 1D chain with two distinct ordering fields. 
We use a Lee-Yang zeros analysis to investigate the effects of negative numbers of invisible states and the complex fields acting 

We found that the loci of Lee-Yang zeros and of the Yang-Lee edges strongly depend on the number of invisible states $r$. 
The area covered by the locus of the Yang-Lee edge increases with $r$ but, if the convention of a positive number of invisible states and real fields are adhered to, the real part of the $z_1$-axis is only crossed at zero temperature.
This extends the original result of  Ising to generalisations of the eponymous model; there is no positive-temperature phase transition.
Nonetheless, the fact that invisible states alter the locus of the partition function zeros opens up new possibilities to achieve this end. 
Interesting results are obtained when either number of invisible states is negative or the external field $h_2$ is complex. Both these cases are shown to deliver the possibility for the system to undergo phase transition at positive temperature. 
Ref.~\cite{Theo08} contains a review of 1D lattice models with entropic stabilities, showing  that the temperature at which the energy cost of producing a domain wall is balanced by gain of entropy matches the temperature at which the transitions occur, thus asking the question if all phase transitions in 1D are driven by the formation of domain walls.
Here, we have established that 1D phase transitions are achievable by negative numbers of invisible states and by complex magnetic fields. These would appear to be outside the domain-wall criterion.

The reasons why our results are not governed by the rigorous theorems \cite{Landau,Ruelle1969} are twofold.
Firstly, the addition of a negative number of invisible states can be interpreted as a sort of indirect or artificial  ordering mechanism, whereby the entropy is decreased by an amount sufficient to bypass the no-go theorems. This suggests that the introduction of a more physically direct  ordering mechanism might overcome the no-go theorems in a new manner. 
The second way around the  no-go theorems is the promotion of magnetic fields to complex values.
Fields acting on conventional states have been shown to be mappable to quantum coherence time.
This suggests that allowing $h_2$ to be complex may similarly endow  our classical system with an element of quantum properties. 
Finally,  according to the equivalent Hamiltonian (\ref{Href}), a negative number of invisible states plays the same role as a complex chemical potential, suggestive of a possible link between both bypass mechanisms.

\section*{Acknowledgements}
We would like to thank Vahan Hovhannisyan and Nerses Ananikian for fruitful discussions. This work was supported in part by FP7 EU IRSES projects No. 295302 "Statistical Physics in Diverse Realizations", No. 612707 "Dynamics of and in Complex Systems".

\newpage

\appendix

\section{Fisher zeros}
\label{Fisher}

In Ref.~\cite{Sarkanych2017}, we reported on an analysis of the Fisher zeros in the complex temperature plane\cite{Fisher1965}. 
Here we present a brief version of that study because 
(i) it shows how we extract zero field critical points needed for Section \ref{IV} and 
(ii)  it provides information on phase transition strength.

Fisher zeros are usually considered at the critical value of the external field.  
For the spontaneous phase transition the critical value of the field is $h_1=0$ corresponding to $z_1=1$. 
In this case one of the roots of the polynomial (\ref{eigmain}) becomes $\lambda=y-1$ so that the polynomial (\ref{eigmain}) has only three different eigenvalues.
Fisher zeros can be obtained from the condition that (at least) two eigenvalues of the transfer matrix are largest by modulus \cite{Fisher1980}. 
This approach allows to obtain zeros of the finite-size system and thus use the finite-size scaling technique (FSS) for the Fisher zeros coordinates. 
In the thermodynamic limit the line of zeros crosses the real axis at the transition point. 

\subsection{Critical temperature}
\label{III21}
Fixing $z_1=z_2=1$ in Eq.~(\ref{largeeq}) we arrive at the equation for the coordinates of the partition function zeros in the complex $y-$plane at given pair of $(q,r)$. 
It is most convenient to display Fisher zeros in the complex $t=y^{-1}$ plane. 
In this case they form closed curves around the origin $t=0$ (rather than the ``run-away'' behaviour at $y\to \infty$). 
The infinite region $0\leq T<\infty$ corresponds to the section $0\leq t\leq 1$. 
In Fig.~\ref{fzeros}(a) we plot coordinates of the zeros for $q=2$ and $r=0,1,2,5$ at fixed $N=128$. The case $r=0$ recovers results for the 1D Ising model. 
For $q=2, r=0$ zeros lie on the imaginary axis. 
With increasing $N$, the zero closest to the real axis closes in and with $N\to \infty$ it crosses the real axis at $t=0$ ($T=0$), implying again that there is only a zero temperature phase transition for the 1D systems \cite{Landau,Ruelle1969}. 
\begin{figure}[t]\begin{center}
		\includegraphics[width=0.3\paperwidth]{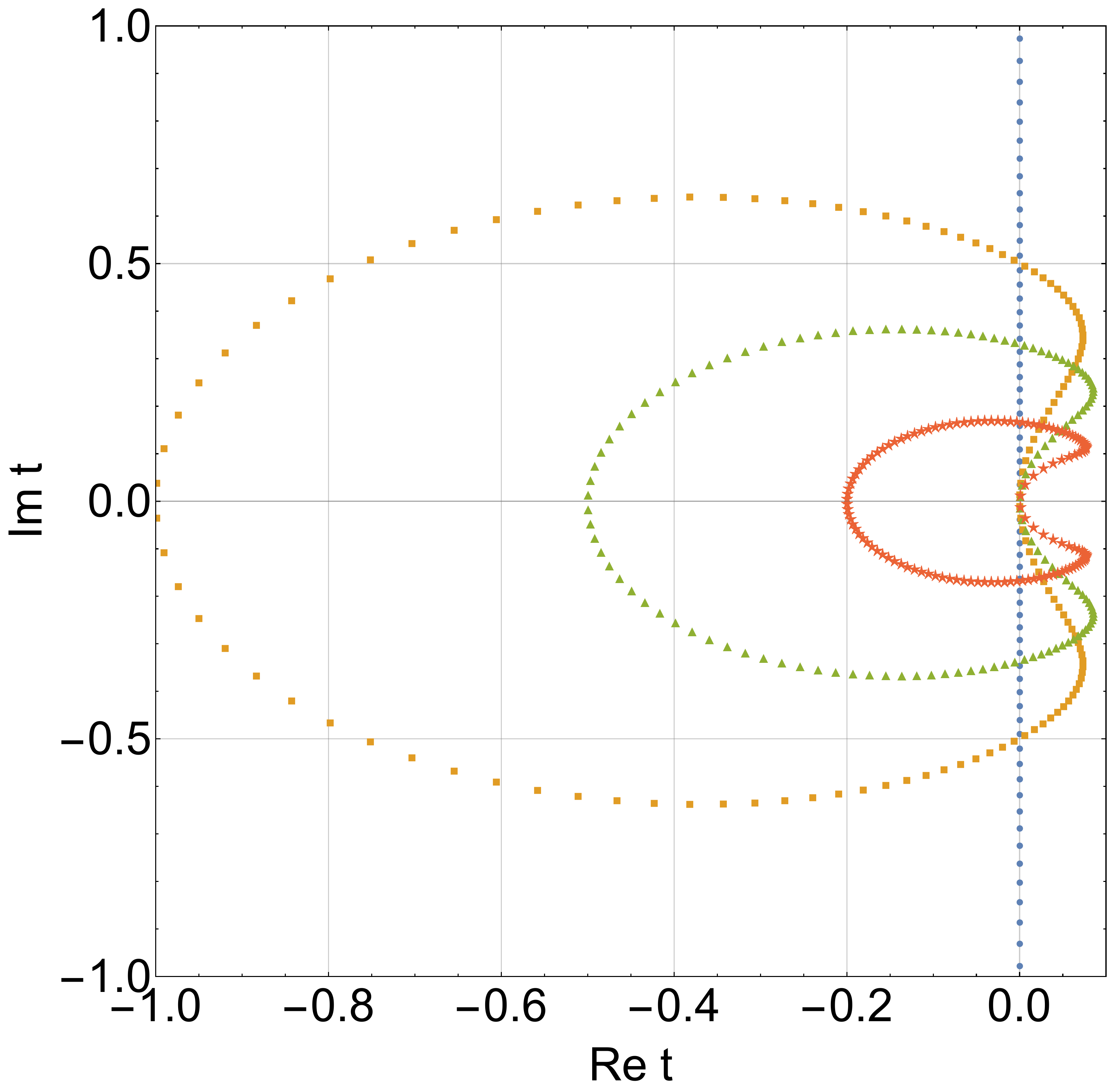} \includegraphics[width=0.3\paperwidth]{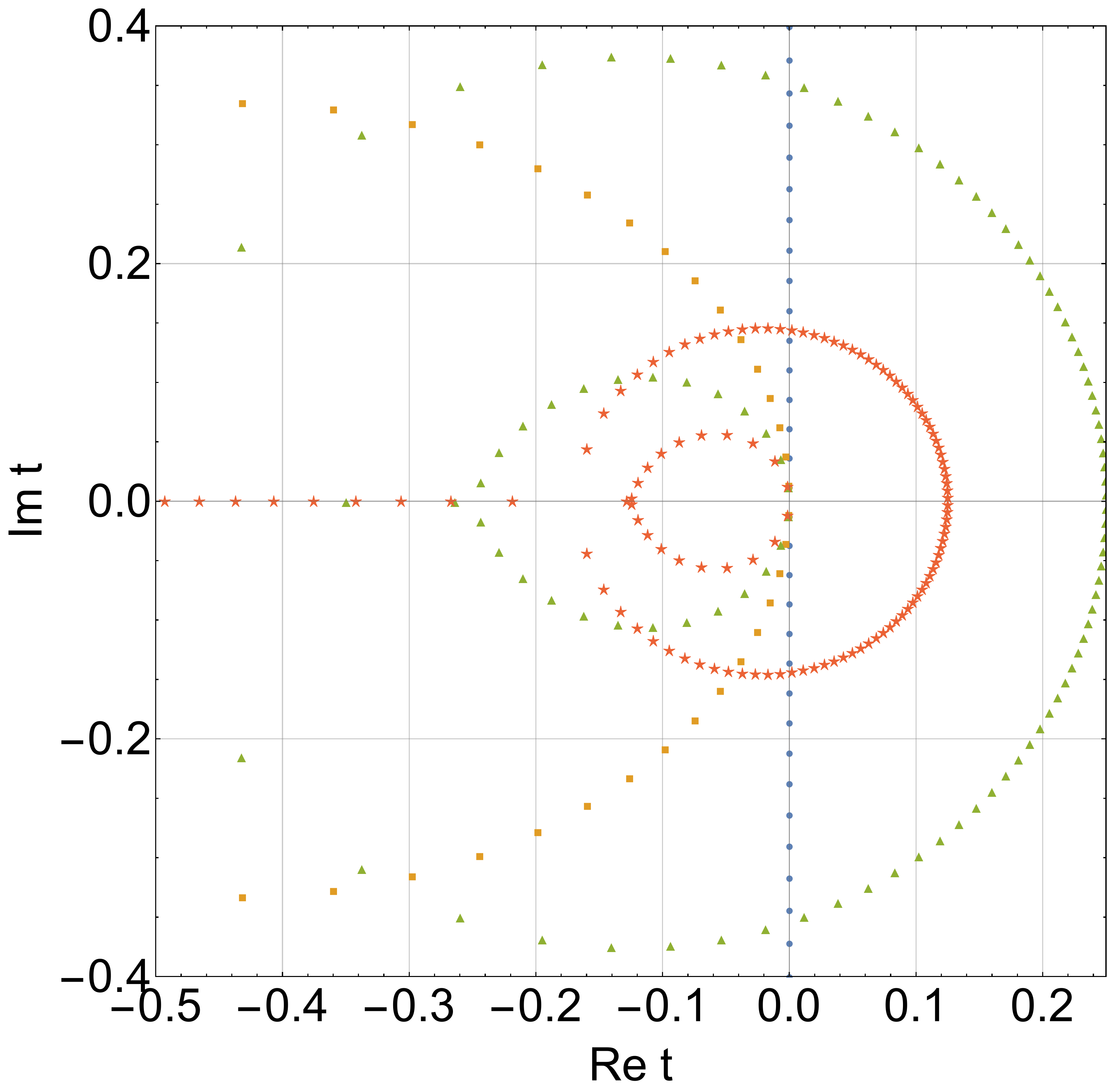}\\
		\quad(a)\hspace*{0.28\paperwidth}(b)
		\caption{ Fisher zeros of the $(2,r)-$state Potts model in $t=y^{-1}=e^{-\beta}$ plane for system size $N=128$ with a) $r=0$(blue), $1$(yellow), $2$(green), $5$(red) and b)  $r=0$(blue), $-2$ (yellow), $-5$ (green), $-7$ (red). \label{fzeros}}
	\end{center}\end{figure}
	
As one can see from Fig.~\ref{fzeros}(a), the presence of invisible states changes the locus of Fisher zeros. 
Now the zeros have both real and imaginary parts and in addition one more crossing point of the real $t-$axis appears. 
However this crossing point is located in the unphysical region $t<0$ (complex values of $T$). 
$t=0$ remains the point where the Fisher zeros approach the real axis and this confirms that the phase transition in the 1D Potts model is not changed by the presence of the invisible states.
The critical exponents of this transition are further discussed in \ref{III22}.

We extend these considerations and show in Fig.~\ref{fzeros}(b) the locus of zeros for the $(2,r)-$state Potts model with negative values of $r$.	\footnote{We do not show in the plot some points in the region $t>1$ that correspond to negative temperatures.}
	
As we can see from the figure, the locus of Fisher zeros in case $r=-5, -7$ intersect the real axis at the value $t_c=-\frac{1}{q+r-1}$. This means that besides the ordinary zero temperature phase transition we observe finite temperature phase transition in 1D model. 
The equivalent representation of the Potts model with invisible states through Eq.~(\ref{Href}) indicates that the chemical potential is $\mu=-T\log r$.
Therefore  the negative number of invisible states is equivalent to a model with complex chemical potential.
Again, via the aforementioned relation  between the complex external field and decoherence time \cite{Wei2012}, this gives a connection to the behaviour of quantum systems.

\subsection{Critical exponents}
\label{III22}
With the coordinates of the Fisher zeros to hand it is possible to obtain values of critical exponents. 
The first method to use is the scaling of the zero closest to the critical point. 
The corresponding scaling law has the form
\begin{eqnarray}
\mbox{Re}\, t=t_c+A\cdot N^{-\Lambda}\\
\mbox{Im}\, t=B\cdot N^{-1/\nu},
\end{eqnarray} 
where $t_c$ is the critical temperature, $\nu$ is the correlation length critical exponent, $\Lambda$ is so-called shift exponent and $N$ is the system size \cite{Itzykson1983}.

Another approach to analyse the partition function zeros is to use partition function zeros density \cite{JaKe}. This method allows to use not only coordinate of the closest zero but consider zeros density function which scales as
\begin{equation}
G(R)\sim R^{2-\alpha},
\end{equation}
where $R$ is the distance to zero from the critical point $\alpha$ is the specific heat critical exponent.

We have used both techniques  to extract the critical exponents from the coordinates of the zeros. 
Using system sizes from $N=500$ to $N=1000$ with increment $\Delta N=20$ $q=2$ and $r=6$ we obtained $\nu=0.9998(2)$, $\Lambda=1.9997(2)$ and $\alpha=1.002(2)$, which are in a good agreement with the hyperscaling relation $\alpha=2-d\nu$. 
Moreover these values remain close to the values $\nu=1, \alpha=1, \Lambda=2$ with $q$ and $r$ changing.

It is worth mentioning that in the absence of magnetic field exact solution can be obtained and critical exponents are the same as in 1D Ising model $(\nu=1,\, \alpha=1,\,\eta=1,\,\gamma=1,\,\mu=0,\,\beta=0,\,\delta=\infty).$


\section{Duality relations}
\label{Duality}

In the main text we have observed that a positive-temperature phase transition can be induced in a short-range 1D model by rendering the number of invisible states $r$ negative or rendering the magnetic field $h_2$ complex. 
Here we point out a duality relation between these two parameters.

In Eq.~(\ref{eigmain}), the quantities $r$ and $z_2$ appear  only in one term together as a sum. 
We conclude that the magnetic field acting on the invisible states plays the same role as additional invisible states. 
In this way $r+z_2-1$ can be treated as a temperature-dependent number of invisible states. 
For this reason,  in this Appendix $z_2$ is included in $r$ and is never shown explicitly. 

Duality means that under a certain unitary transformation $\mathbf{S}$, the transfer matrix changes according to the rule \cite{Kramers1941}
\begin{equation}
\label{dual1}
\mathbf{ST}(y,z_1)\mathbf{S}^{-1}=\alpha \mathbf{T}^T(y^D, z_1^D),
\end{equation}
where $y^D=y^D(y,z_1),z_1^D=z_1^D(y,z_1)$ denote variables dual to $y,z_1$, and $\mathbf{T}^T$ is the transposed transfer matrix. 
Eq.~({\ref{dual1}}) can be rewritten in terms of the eigenvalues
\begin{equation}
\lambda(y^D,z_1^D)=\frac1{\alpha(y,z_1)}\lambda(y,z_1)\label{dualstart}.
\end{equation}

Eq.~(\ref{dualstart}) is useful when explicit expressions for the eigenvalues are known. 
But in our case solving the third-order equation (\ref{eigmain}) results in having cumbersome  expressions for each $\lambda$ and thus Eq.~(\ref{dualstart}) will be hard to handle. 
Instead, let us derive relations for the coefficients in the third-order polynomial in the left hand side of Eq.~(\ref{eigmain}). 
In the most general case the equation reads
\begin{equation}
\label{dualgeneraleq}
\lambda^3+A_1(y,z_1)\lambda^2+A_2(y,z_1)\lambda+A_3(y,z_1)=0.
\end{equation}
Eq.~(\ref{dualgeneraleq}) holds for both ordinary and dual variables. Substituting  (\ref{dualstart}) into (\ref{dualgeneraleq}) we get the following transformation rules for the coefficients $A_1, A_2, A_3$:
\begin{gather}
A_1(y^D,z_1^D)=\frac1\alpha A_1(y,z_1),\nonumber\\
A_2(y^D,z_1^D)=\frac1{\alpha^2} A_2(y,z_1), \label{15}\\
A_3(y^D,z_1^D)=\frac1{\alpha^3} A_3(y,z_1)\nonumber.
\end{gather}

Using the first two equations of (\ref{15}) and Eq.~(\ref{dualstart}) with $\lambda=y-1$ one  recovers expressions for the dual variables:
\begin{gather}
\nonumber \alpha=\frac{(y-1) (q (z_1-1)+(r-1) z_1+1)}{q^2+q (2 r-1)+(r-2) r}, \\
\label{dual} y^D=\frac{(q+r-1) (q+r+z_1-1)}{q (z_1-1)+(r-1) z_1+1},\\
\nonumber z_1^D=\frac{q^2+q (2 r+y-2)+r^2-2 r-y+1}{(y-1) (q+r-1)}.
\end{gather}
These expressions allow us to substitute temperature by field and vice versa without changing the behaviour of the system. Substituting $r=0$ into (\ref{dual}) one  recovers the duality relations for the ordinary 1D Potts model obtained in \cite{Glumac1994}.

\section{Yang-Lee edge singularity exponent}
\label{Singular}
In this appendix we show that the value $\sigma=\frac12$ remain unchanged by introdusing invisible states. To do so we closely follow the method developed in Refs.~\cite{Hovhannisyan2009,Ghulghazaryan2007}.

With increase of system size, Lee-Yang zeros $z_1=|z_1|e^{i\theta}$ terminate in the complex plane at the Yang-Lee edge $z_1^e=|z_1^e|e^{i\theta_e}$. 
Their density $g(z_1)$ in the vicinity of $z_1^e$ is governed by the edge singularity exponent $\sigma$~\cite{KoGr71}:
\begin{equation}
g(\theta)\propto |\theta-\theta_e|^\sigma \, .
\end{equation}
In circumstances where, for a given value of $T$, the zeros are located on curves (the so-called singular line \cite{APPENDIXsingline} as opposed to two dimensional regions \cite{JaKe}), the function $g$ can be written for a fixed $|z_1|$ keeping dependency of the phase $\theta$ only.
The exponent $\sigma$, like the other critical exponents, is characteristic of a given universality class. 
For the 1D Ising and $q-$state Potts models its exact value is $\sigma=-\frac12$ \cite{LY,KoGr71,Ghulghazaryan2007}.
Another known exact value for the $\sigma-$exponent has been obtained for the spherical model, where $\sigma=\frac12$ independently of the type of interaction (short- or long-range) and space dimensionality \cite{Kurtze1978}.

The density of the partition function zeros in the region $(\phi,\phi+\Delta\phi)$ is proportional to the number of zeros in this region divided by the length of the part of the cord these zeros occupy. 
Since for each $\phi$ there is a certain zero, than the number of zeros in the region $(\phi,\phi+\Delta\phi)$ is proportional to $\Delta\phi$. 
Therefore, the density can be written as 
\begin{equation}
\label{dens0}
\tilde g(\phi,\phi+\Delta\phi)\propto \dfrac{\Delta\phi}{\int_{\phi}^{\phi+\Delta\phi}{\sqrt{(\frac{\partial \mathrm{Re} z_1}{\partial \phi})^2+(\frac{\partial \mathrm{Im} z_1}{\partial \phi})^2}\,d\phi}}\,.
\end{equation}
In the thermodynamic limit zeros form continuous curve with density at the point $\phi$ given by  
\begin{equation}
\label{dens1}
\tilde g(\phi,\phi+d\phi)=g(\phi)\propto \cfrac{1}{\sqrt{(\frac{\partial \mathrm{Re} z_1}{\partial \phi})^2+(\frac{\partial \mathrm{Im} z_1}{\partial \phi})^2}}\,.
\end{equation}
In the vicinity of the edge  $\theta_e$ (which corresponds to $\phi=0$) coordinates of zeros can be expanded into the Taylor series
\begin{equation}
\label{expansion}
z_1(\phi)\approx z_1^e+\cfrac{\partial^2 z_1(0)}{\partial \phi^2}\phi^2+ \ldots \quad,
\end{equation}
The linear term is absent since Eq.~(\ref{largeeq}) is an even function of $\phi$. 
Substituting the expansion (\ref{expansion}) into Eq.~(\ref{dens1}) we obtain a simple relation between the density function $g(\phi)$ and the phase $\phi$:
\begin{equation}
\label{dens2}
g(\phi)\propto |\phi|^{-1} \,.
\end{equation}
In the thermodynamic limit, close to the edge point, the phase and coordinates of zeros are connected through
\begin{equation}
\label{dens3}
\theta-\theta_e \propto |z_1-z_1^e| \,.
\end{equation}
Using expansion (\ref{expansion}) in the right-hand side of the Eq.~(\ref{dens3}) we arrive at 
\begin{equation}
\label{dens4}
\phi^2 \propto (\theta-\theta_e) \,.
\end{equation} 
Relation (\ref{dens2}) together with Eq.~(\ref{dens4}) lead to the power-law behaviour of the density of zeros as a function of their phase close to the edge point   
\begin{equation}
g(\theta)\propto |\theta-\theta_c|^{-1/2} \,.
\end{equation}
Thus the Yang-Lee edge singularity exponent is $\sigma=-\frac12$. This value follows immediately from the symmetry of zeros under the substitution $\phi\to -\phi$, which is observed for the models considered in Refs.~\cite{Glumac1994,Ghulghazaryan2007,Hovhannisyan2009,Ananikian2013}.


\section*{References}

\begin{thebibliography}{10}
	\expandafter\ifx\csname url\endcsname\relax
	\def\url#1{{\tt #1}}\fi
	\expandafter\ifx\csname urlprefix\endcsname\relax\def\urlprefix{URL }\fi
	\providecommand{\eprint}[2][]{\url{#2}}
	
	
	
\bibitem{Ising}
Ising, E. (1925). Beitrag zur theorie des ferromagnetismus. {\em Zeitschrift f\"ur Physik}, \textbf{31}(1), 253.
	
\bibitem{Ising17}
Ising, T., Folk, R., Kenna, R., Berche, B., \& Holovatch, Y. (2017). The Fate of Ernst Ising and the Fate of his Model. {\em Journal of Physical Studies}, \textbf{21}(3), 3009.

\bibitem{Rowlinson} 
Kipnis, Y. R., Kipnis, A. Y., \& Rowlinson, J. S. (1994). Van der Waals and molecular science. {\em Oxford University Press}.

\bibitem{Landau}
Lifshitz, E. M., \& Pitaevskii, L. P. (2013). Statistical physics: theory of the condensed state (Vol. 9). {\em Elsevier}.
	
\bibitem{Hove}
Van Hove, L. (1950). Sur l'int\'egrale de configuration pour les syst\'emes de particules \`a une dimension. {\em Physica}, \textbf{16}(2), 137.

\bibitem{Ruelle1969}	
Ruelle, D. (1968). Statistical mechanics of a one-dimensional lattice gas. {\em Communications in Mathematical Physics}, \textbf{9}(4), 267;
Ruelle, D. (1999). Statistical mechanics: Rigorous results. {\em World Scientific}.
		
\bibitem{CuSa}
Cuesta, J. A., \& S\'anchez, A. (2004). General non-existence theorem for phase transitions in one-dimensional systems with short range interactions, and physical examples of such transitions. \textit{Journal of statistical physics}, \textbf{115}(3-4), 869.

\bibitem{Theo08}
Theodorakopoulos, N. (2006). Phase transitions in one dimension: Are they all driven by domain walls? \textit{Physica D: Nonlinear Phenomena}, \textbf{216}(1), 185.

\bibitem{Anderson}
Anderson, P. W., \& Yuval, G. (1969). Exact results in the Kondo problem: equivalence to a classical one-dimensional Coulomb gas.\textit{ Physical Review Letters}, \textbf{23}(2), 89;
Yuval, G., \& Anderson, P. W. (1970). Exact results for the Kondo problem: One-body theory and extension to finite temperature. \textit{Physical Review B}, \textbf{1}(4), 1522;
Anderson, P. W., Yuval, G., \& Hamann, D. R. (1970). Exact results in the Kondo problem. II. Scaling theory, qualitatively correct solution, and some new results on one-dimensional classical statistical models. \textit{Physical Review B}, \textbf{1}(11), 4464.

\bibitem{Dyson71}
Dyson, F. J. (1971). An Ising ferromagnet with discontinuous long-range order. \textit{Communications in Mathematical Physics}, \textbf{21}(4), 269.

\bibitem{FrSp82}
Fr\"ohlich, J., \& Spencer, T. (1982). The phase transition in the one-dimensional Ising model with $1/r^2$ interaction energy.\textit{ Communications in Mathematical Physics}, \textbf{84}(1), 87.

\bibitem{Evans}
Evans, M. R. (2000). Phase transitions in one-dimensional nonequilibrium systems. \textit{Brazilian Journal of Physics}, \textbf{30}(1), 42.

\bibitem{Sarkanych2017}
Sarkanych, P., Holovatch, Y., \& Kenna, R. (2017). Exact solution of a classical short-range spin model with a phase transition in one dimension: the Potts model with invisible states. \textit{Physics Letters A}, \textbf{381}(41), 3589.

\bibitem{LY}
Yang, C. N., \& Lee, T. D. (1952). Statistical theory of equations of state and phase transitions. I. Theory of condensation. \textit{Physical Review}, \textbf{87}(3), 404;
Lee, T. D., \& Yang, C. N. (1952). Statistical theory of equations of state and phase transitions. II. Lattice gas and Ising model. \textit{Physical Review}, \textbf{87}(3), 410.
	
\bibitem{fund}
Wu, F. Y. (2008). Professor C.N. Yang and statistical mechanics. \textit{International Journal of Modern Physics B}, \textbf{22}(12), 1899.
	
\bibitem{Tamura2010}
Tamura, R., Tanaka, S., \& Kawashima, N. (2010). Phase transition in Potts model with invisible states. \textit{Progress of theoretical physics}, \textbf{124}(2), 381.
	
\bibitem{Tanaka2011}
Tanaka, S., \& Tamura, R. (2011). Dynamical properties of Potts model with invisible states.  \textit{Journal of Physics: Conference Series}, \textbf{320}(1), 012025.

\bibitem{Potts1952}
Potts, R. B. (1952). Some generalized order-disorder transformations. \textit{Mathematical proceedings of the cambridge philosophical society}, \textbf{48}(1), 106.
	
\bibitem{Krasnytska2016}
Krasnytska, M., Sarkanych, P., Berche, B., Holovatch, Y., \& Kenna, R. (2016). Marginal dimensions of the Potts model with invisible states. \textit{Journal of Physics A: Mathematical and Theoretical},\textbf{ 49}(25), 255001.

\bibitem{Wei2012}
Wei, B. B., \& Liu, R. B. (2012). Lee-Yang zeros and critical times in decoherence of a probe spin coupled to a bath. \textit{Physical review letters}, \textbf{109}(18), 185701.

\bibitem{Peng2015}
Peng, X., Zhou, H., Wei, B. B., Cui, J., Du, J., \& Liu, R. B. (2015). Experimental observation of Lee-Yang zeros. \textit{Physical Review Letters}, \textbf{114}(1), 010601.

\bibitem{Bena2005}
Bena, I., Droz, M., \& Lipowski, A. (2005). Statistical mechanics of equilibrium and nonequilibrium phase transitions: the Yang--Lee formalism. \textit{International Journal of Modern Physics B}, \textbf{19}(29), 4269.
	
\bibitem{TiCh17}
Timonin, P. N., \& Chitov, G. Y. (2017). Infinite cascades of phase transitions in the classical Ising chain. \textit{Physical Review E}, \textbf{96}(6), 062123.	
	
\bibitem{Seth2017}
Seth, S. (2016). Combinatorial approach to exactly solve the 1D Ising model. \textit{European Journal of Physics}, \textbf{38}(1), 015104.

\bibitem{Onsager}
Onsager, L. (1944). Crystal statistics. A two-dimensional model with an order-disorder transition. \textit{Physical Review}, \textbf{65}(3-4), 117.

\bibitem{LeLe16}
Lee, J. U., Lee, S., Ryoo, J. H., Kang, S., Kim, T. Y., Kim, P., Park, C.-H., Park, J.-G. \& Cheong, H. (2016). Ising-Type magnetic ordering in atomically thin $FePS_3$. \textit{Nano letters}, \textbf{16}(12), 7433.
	
\bibitem{Press}
Press release: Onsager's predicted magnetic phase transition in 2D materials observed at last
http://nanotechweb.org/cws/article/yournews/67403

\bibitem{Suzuki1976}
Suzuki, M. (1976). 
Relationship between $d$-dimensional quantal spin systems and $(d+1)$-dimensional Ising systems: Equivalence, critical exponents and systematic approximants of the partition function and spin correlations. 
{\em Progress of theoretical physics}, {\bf 56}(5), 1454.

\bibitem{Rice1993}
Rice, T. M., Gopalan, S., \& Sigrist, M. (1993). Superconductivity, spin gaps and Luttinger liquids in a class of cuprates. \textit{EPL (Europhysics Letters)}, \textbf{23}(6), 445.
	
\bibitem{Rice1997}
Rice, T. M. (1996). $tJ$ ladders and cuprate ladder compounds.\textit{ Zeitschrift f\"ur Physik B Condensed Matter}, \textbf{103}(2), 165.
	
\bibitem{Azuma1994}
Azuma, M., Hiroi, Z., Takano, M., Ishida, K., \& Kitaoka, Y. (1994). Observation of a Spin Gap in $Sr Cu_2 O_3$ Comprising Spin-1/2 Quasi-1D Two-Leg Ladders. \textit{Physical review letters}, \textbf{73}(25), 3463.
	
\bibitem{Roger1983}
Roger, M., Hetherington, J. H., \& Delrieu, J. M. (1983). Magnetism in solid He3. \textit{Reviews of Modern Physics},\textbf{ 55}(1), 1.
	
\bibitem{Ananikian}
Ananikian, N. S., Stre\v{c}ka, J., Hovhannisyan, V. (2014). 
Magnetization plateaus of an exactly solvable spin-1 Ising--Heisenberg diamond chain. 
	\textit{Solid State Communications}, \textbf{194}, 48.
	
\bibitem{Hovhannisyan2009}
Hovhannisyan, V. V., Ghulghazaryan, R. G., \& Ananikian, N. S. (2009). The partition function zeros of the anisotropic Ising model with multisite interactions on a zigzag ladder. \textit{Physica A: Statistical Mechanics and its Applications}, \textbf{388}(8), 1479.
	 
\bibitem{Badasyan2010}
Badasyan, A. V., Giacometti, A., Mamasakhlisov, Y. S., Morozov, V. F., \& Benight, A. S. (2010). Microscopic formulation of the Zimm-Bragg model for the helix-coil transition. \textit{Physical Review E}, \textbf{81}(2), 021921.
	 	
\bibitem{Badasyan2011}
Badasyan, A. V., Tonoyan, S. A., Mamasakhlisov, Y. S., Giacometti, A., Benight, A. S., \& Morozov, V. F. (2011). Competition for hydrogen-bond formation in the helix-coil transition and protein folding. \textit{Physical Review E}, \textbf{83}(5), 051903.
	 	
\bibitem{Johnston2013}
Johnston, D. A., \& Ranasinghe, R. P. K. C. M. (2013). Potts models with (17) invisible states on thin graphs. \textit{Journal of Physics A: Mathematical and Theoretical}, \textbf{46}(22), 225001.
	
\bibitem{Ananikian2013}
Ananikian, N., Izmailyan, N. S., Johnston, D. A., Kenna, R., \& Ranasinghe, R. P. K. C. M. (2013). Potts models with invisible states on general Bethe lattices. \textit{Journal of Physics A: Mathematical and Theoretical},\textbf{ 46}(38), 385002.

\bibitem{Baxter1982}
Baxter, R. J. (2016). Exactly solved models in statistical mechanics. \textit{Elsevier}.

\bibitem{Katsura1972}
Katsura, S., \& Ohminami, M. (1972). Distribution of zeros of the partition function for the one dimensional Ising models. \textit{Journal of Physics A: General Physics}, \textbf{5}(1), 95.

\bibitem{Kim2000}
Kim, S. Y., \& Creswick, R. J. (2000). Exact results for the zeros of the partition function of the Potts model on finite lattices.\textit{ Physica A: Statistical Mechanics and its Applications}, \textbf{281}(1-4), 252.

\bibitem{Shrock1997}
Shrock, R., \& Tsai, S. H. (1997). Complex-temperature phase diagrams of one-dimensional spin models with next-nearest-neighbor couplings. \textit{Physical Review E}, \textbf{55}(5), 5184.	
	
\bibitem{Fisher1965}
Fisher, M. E. (1965).  The nature of critical points.  In ed. Britten, W.E.,
Lectures in theoretical physics, v. 7C, p. 1--159. \textit{University of Colorado Press, Boulder, Colorado, USA}.
	
\bibitem{Fisher1980}
Fisher, M. E. (1980). Yang-Lee edge behavior in one-dimensional systems. \textit{Progress of Theoretical Physics Supplement}, \textbf{69}, 14.
	
\bibitem{Ghulghazaryan2007}
Ghulghazaryan, R. G., Sargsyan, K. G., \& Ananikian, N. S. (2007). Partition function zeros of the one-dimensional Blume-Capel model in transfer matrix formalism. \textit{Physical Review E}, \textbf{76}(2), 021104.
	
\bibitem{Fi78}
Fisher, M. E. (1978). Yang-Lee edge singularity and $\phi^3$ field theory. \textit{Physical Review Letters}, \textbf{40}(25), 1610.

	
\bibitem{Glumac1994}
Glumac, Z., \& Uzelac, K. (1994). The partition function zeros in the one-dimensional q-state Potts model. \textit{Journal of Physics A: Mathematical and General}, \textbf{27}(23), 7709.
	
\bibitem{KoGr71}
Kortman, P. J., \& Griffiths, R. B. (1971). Density of zeros on the Lee-Yang circle for two Ising ferromagnets. \textit{Physical Review Letters},\textbf{ 27}(21), 1439.
	
\bibitem{Wang}
Wang, X. (2012). Physical examples of phase transition in one-dimensional systems with short range
interaction,\\
\url{http://guava.physics.uiuc.edu/\textasciitilde nigel/courses/563/Essays\_2012/PDF/wang.pdf}
	
\bibitem{Kim2001}
Kim, S. Y., \& Creswick, R. J. (2001). Density of states, Potts zeros, and Fisher zeros of the Q-state Potts model for continuous Q. \textit{Physical Review E}, \textbf{63}(6), 066107.
	
\bibitem{Ghulghazaryan2003}
Ghulghazaryan, R. G., \& Ananikian, N. S. (2003). Partition function zeros of the one-dimensional Potts model: the recursive method. \textit{Journal of Physics A: Mathematical and General}, \textbf{36}(23), 6297.
	
\bibitem{Glumac2002}
Glumac, Z., \& Uzelac, K. (2002). Complex-q zeros of the partition function of the Potts model with long-range interactions. \textit{Physica A: Statistical Mechanics and its Applications}, \textbf{310}(1-2), 91.
	
\bibitem{q1}
Kasteleyn, P. W., \& Fortuin, C. M. (1969). Phase transitions in lattice systems with random local properties. \textit{Journal of the Physical Society of Japan Supplement}, \textbf{26}, 11.
	
\bibitem{q0}
Jacobsen, J. L., Salas, J., \& Sokal, A. D. (2005). Spanning forests and the q-state Potts model in the limit $q\to 0$. \textit{Journal of Statistical Physics}, \textbf{119}(5-6), 1153;
Majumdar, S. N., \& Dhar, D. (1992). Equivalence between the Abelian sandpile model and the $q\to 0$ limit of the Potts model. \textit{Physica A: Statistical Mechanics and its Applications}, \textbf{185}(1-4), 129; 
Harris, A. B., \& Lubensky, T. C. (1987). Potts-model formulation of the random resistor network. \textit{Physical Review B}, \textbf{35}(13), 6987.

\bibitem{Bonati2010}
Bonati, C., \& D'Elia, M. (2010). Three-dimensional, three-state Potts model in a negative external field. \textit{Physical Review D}, \textbf{82}(11), 114515.

\bibitem{Dalmazi2008}
Dalmazi, D., \& S\`a, F. L. (2008). The Yang--Lee edge singularity in spin models on connected and non-connected rings. \textit{Journal of Physics A: Mathematical and Theoretical}, \textbf{41}(50), 505002.	

\bibitem{Itzykson1983}
Itzykson, C., Pearson, R. B., \& Zuber, J. B. (1983). Distribution of zeros in Ising and gauge models. \textit{Nuclear Physics B}, \textbf{220}(4), 415.

\bibitem{JaKe}
Janke, W., \& Kenna, R. (2001). The strength of first and second order phase transitions from partition function zeroes. \textit{Journal of Statistical Physics}, \textbf{102}(5-6), 1211;
Janke, W., Johnston, D. A., \& Kenna, R. (2004). Phase transition strength through densities of general distributions of zeroes. \textit{Nuclear Physics B}, \textbf{682}(3), 618.

\bibitem{Kramers1941}
Kramers, H. A., \& Wannier, G. H. (1941). Statistics of the two-dimensional ferromagnet. \textit{Physical Review}, \textbf{60}(3), 252.
		
\bibitem{APPENDIXsingline}
Abe, R. (1967). Logarithmic singularity of specific heat near the transition point in the Ising model. \textit{Progress of Theoretical Physics}, \textbf{37}(6), 1070; 
Abe, R. (1967). Note on the critical behavior of Ising ferromagnets. \textit{Progress of Theoretical Physics}, \textbf{38}(1), 72; 
Abe, R. (1967). Singularity of specific heat in the second order phase transition. \textit{Progress of Theoretical Physics}, \textbf{38}(2), 322; 
Abe, R. (1967). Critical behavior of pair correlation function in Ising Ferromagnets. \textit{Progress of Theoretical Physics}, \textbf{38}(3), 568.

\bibitem{Kurtze1978}
Kurtze, D. A., \& Fisher, M. E. (1978). The Yang-Lee edge singularity in spherical models. \textit{Journal of Statistical Physics}, \textbf{19}(3), 205.

\end{thebibliography}
\end{document}